\documentclass[%
 reprint,
 amsmath,amssymb,
 aps,
]{revtex4-2}
\usepackage{comment}
\usepackage{graphicx}
\usepackage{dcolumn}
\usepackage{bm}
\usepackage{hyperref}
\usepackage{cleveref}

\usepackage{nicefrac}
\usepackage{physics}
\PassOptionsToPackage{square,numbers}{natbib}
\usepackage[dvipsnames]{xcolor}
\usepackage{amsmath,amsfonts,amsthm,amssymb}
\usepackage{booktabs} 
\usepackage{comment} 
\usepackage[utf8]{inputenc} 
\usepackage{graphicx}
\usepackage{braket}
\usepackage{float,tcolorbox}
\usepackage{subcaption}
\usepackage[mode=buildnew]{standalone}

\usepackage{placeins}

\usepackage{tikz}
\usepackage[mode=buildnew]{standalone}

\usepackage{blindtext} 
\usepackage{ragged2e}
\usepackage{float}
\usepackage{scrextend}
\usepackage{footmisc}
\usepackage{booktabs}
\usepackage{multirow}

\usepackage{tikz}
\usetikzlibrary{arrows.meta, positioning}

\tikzset{
  mystep/.style={
    rectangle,
    minimum width=\dimexpr\columnwidth-1cm\relax,
    text centered,
    inner sep=5pt
  },
  myarrow/.style={
    thick,
    ->,
    >=stealth
  }
}

\usepackage{blindtext} 
\usepackage{ragged2e}

\newcommand{\gcsqrtzero}{\operatorname{U}_{\operatorname{CH}}^{01}}
\newcommand{\gcsqrtone}{\operatorname{U}_{\operatorname{CH}}^{12}}
\newcommand{\gcxzero}{\operatorname{U}_{\operatorname{CX}}^{01}}
\newcommand{\gcxone}{\operatorname{U}_{\operatorname{CX}}^{12}}
\newcommand{\ucrzeroone}{\operatorname{U}_{\operatorname{CR}}^{01}}
\newcommand{\ucronetwo}{\operatorname{U}_{\operatorname{CR}}^{12}}
\newcommand{\ucr}{\operatorname{U}_{\operatorname{CR}}}

\begin{document}

\preprint{APS/123-QED}

\title{Generalized cross-resonance scheme for maximally-entangling two-qutrit gates}%

\author{Yash Saxena$^{1,2}$}
 \email{ysaxena@gapp.nthu.edu.tw}
 \author{Tharrmashastha SAPV$^{2,3}$}
 \email{tharrmashasthav@iiitd.ac.in}
\author{Sagnik Chatterjee$^{2,3,4}$}
\email{sagnikc@iiitd.ac.in}
\author{Ray-Kuang Lee$^{1,5,6}$} 
\email{rklee@ee.nthu.edu.tw}

\affiliation{
$^1$Department of Physics, National Tsing Hua University, Hsinchu City, Taiwan\\
$^2$Center for Quantum Technologies, IIIT-Delhi, New Delhi, India\\
$^3$Department of Computer Science, IIIT-Delhi, New Delhi, India\\
$^4$School of Technology and Computer Science, TIFR Mumbai, India\\
$^5$Institute of Photonics Technologies, National Tsing Hua University, Hsinchu 30013, Taiwan\\
$^6$Center for Quantum Science and Technology, Hsinchu, 30013, Taiwan
 }
\begin{abstract}
To utilize higher-dimensional quantum systems, in this Letter, we derive a  generalized cross-resonance (GCR) scheme for realizing maximally entangling two-qutrit gates on fixed-frequency transmons  beyond the 0-1 subspace. Our two-qutrit gates, namely, $\ucrzeroone$ and $\ucronetwo$, acting on the $0{\text -}1$ and $1{\text -}2$ energy transitions of transmons, respectively,  \emph{directly} allow for entanglement on the $1{\text -}2$ levels. Unlike the known works, our gate is parametric in nature, enabling us to construct multiple entangling gates of interest. By performing simulations in Qiskit, we demonstrate two-qutrit generalized controlled-$X$ ($\gcxzero$ and $\gcxone$) and controlled-$H$ ($\gcsqrtzero$ and $\gcsqrtone$) gates, which are instances of the proposed $\ucr$ gates, with reported gate fidelities of $86.14\%~(99.73\%),~84.6\%~(97.88\%),~92.35\%~(99.39\%)$, and $91.99\%~(98.99\%)$, respectively with (and without) noise. We also reveal a two-qutrit Bell state with a fidelity of $99.06 \pm 0.01\%$, with a complete Bell state preparation in a $\sim514$ ns pulse sequence, which is less than the gate time of the known scheme by cross-Kerr-based entangling gates.

\end{abstract}

\maketitle


Working with multi-level quantum systems instead of qubit systems offers many intrinsic advantages w.r.t. quantum computation and information. Access to larger Hilbert spaces has proved beneficial for a myriad of tasks such as qubit reset~\citep{2021qubitreset,PhysRevA.108.052605,PhysRevApplied.10.044030}, qubit readout~\citep{jurcevic2021demonstration}, fidelity in quantum teleportation~\citep{PhysRevA.105.022610,PhysRevA.100.062311}, randomness generation~\citep{PhysRevLett.120.260401}, entanglement purification~\citep{PhysRevA.98.042309}, construction of qudit quantum error-correcting codes~\citep{PhysRevA.103.042420,uy2024quditbasedquantumerrorcorrectingcodes,tanggara2025simpleconstructionquditfloquet}, and to improve security, robustness, and enhance key rates for QKD protocols~\citep{PhysRevA.69.032313,PhysRevLett.96.090501,PhysRevA.67.012311}. Qudit systems have also been provably shown to exhibit speedup and separations in quantum circuit complexity compared to their classical counterparts~\citep{grilo2024powershallowdepthtoffoliqudit,hsieh2024unconditionallyseparatingnoisymathsfqnc0}. More concretely, it has been observed that fewer qudit gates are required to implement a given unitary operation compared to the qubit case ~\citep{Stroud01112002,PhysRevA.75.022313,PhysRevA.103.032417,PhysRevA.111.032408}. Three-level systems or \textit{qutrits}, specifically, have been used to obtain sublinear depth decompositions of qubit circuits without ancilla~\citep{gokhale2019asymptoticimprovement,baker2020intermediatequtrits}.
This reduction in the depth leads to an improvement of the scalability of quantum circuits primarily by ensuring longer decoherence times~\cite{kais,Miyahara2023}.
\par
Quantum devices consisting of mesoscopic anharmonic oscillators have emerged as the leading architecture in the race for realizing physical quantum systems over the last decade~\cite{arute2019quantum}. Circuit Quantum Electrodynamics (cQED) studies the interaction of anharmonic oscillators constructed from nonlinear superconducting circuit elements such as the Josephson junction, with quantized electromagnetic fields in the microwave-frequency domain~\cite{Blais04,Blais21}. The energy spectrum of cQED devices is governed by macroscopic circuit element parameters, which makes these devices highly configurable. Moreover, superconducting transmons demonstrate large nonlinearity~\cite{Fink2008,Schuster2008,Bishop2009,Reed10}, thereby allowing us to restrict the dynamics of the quantum system precisely to the number of levels we desire, making them particularly favourable for implementing qutrits and other higher-level systems.
\par
Consequently, research intensity on extending cQED devices to three-level (and higher) systems has increased, resulting in a flurry of recent works focused on implementing qutrit-based quantum algorithms on superconducting transmons~\cite{Roy23,Liu23}. However, currently, in order to implement arbitrary qutrit-based quantum algorithms on superconducting qutrits, one needs to design building blocks specifically for transmons that allow us to implement a set of qutrit gates that is \textit{universal} for three-level quantum computation. Exact universality on qutrits requires a discrete set of single-qutrit Hamiltonians and one two-qutrit Hamiltonian~\cite{Vlasov02qudit,brylinski2002mathematics,Brenen05qudit}. While high-fidelity single-qutrit quantum gates have long been realized~\cite{QutritHadamard20,QRB21,CharQRB21,Liu23Performing}, realizations of two-qutrit entangling gates using cross-resonance scheme either suffered from increased crosstalk or were limited to perform a direct entanglement on only the $0{-}1$ subspace of the entire two-qutrit Hilbert space~\cite{Scrambling21,Fischer23Universal}. Researchers addressed these issues by realizing maximally entangling two-qutrit gates on the superconducting transmon architecture using differential AC Stark shift drives with non-tunable and tunable couplings~\cite{Goss2022-nature,LuoExperimental}. However, using small cross-Kerr coupling for entangling gates leads to slow gate operations. 

\par


In this letter, we address the challenges posed due to entanglement confined to $0{-}1$ subspace and slow gate times, by extending the well-established and faster cross-resonance scheme to a generalized qutrit architecture. For this, we theorize and demonstrate, via simulations, two-qutrit gates  on fixed-frequency, fixed-coupling transmons that enable high-fidelity, maximally entangling operations across the entire two-qutrit space. This work provides a viable route toward universal qutrit-based quantum computation using experimentally accessible resources. Moving forward, we demonstrate two-qutrit generalized controlled-$X$ and controlled-$H$ gates, using the cross-resonance framework. The tunability required for effective interaction in the two transmon qutrits with fixed frequencies can be achieved using microwave pulse control~\cite{de2010selective, leek2009using, yamamoto2003demonstration, chow2010optimized}. Our simulation setup consists of fixed-frequency transmons (modelled as qutrits) coupled via a resonator. The individual frequencies of the two qutrits are $ \omega_1/2\pi = 4.9~GHz$ and $ \omega_2/2\pi = 5.5~GHz$. The anharmonicities considered for the simulations are $\delta_1=-400~MHz$ and $\delta_2=-300~MHz$. The qutrits interact with a coupling strength of $J / 2\pi = 2.7~\text{MHz}$, and are individually coupled to the control lines with Rabi strengths $g_1 / 2\pi = 0.62~\text{GHz}$ and $g_2 / 2\pi = 0.64~\text{GHz}$, ensuring reliable gate execution.

By performing simulations using Qiskit Dynamics, we realize high-fidelity two-qutrit generalized controlled-$X$ and controlled-$H$ gates,  namely, $U_{CX}^{01}$, $U_{CX}^{12}$, $U_{CH}^{01}$, and $U_{CH}^{12}$, with and without decoherence noises.
Furthermore, we prepare two-qutrit Bell states with fidelities of $99.06\pm0.01\%$ and $78.73 \pm 0.01\%$ in the noiseless and the noisy settings, respectively. Notably, the fidelity in the noiseless setting is comparable to the current state-of-the-art, and has been achieved using a relatively simpler technique for the current superconducting quantum hardware. 
\par
Two-qutrit interactions for the realization of entangling gates are controllable entirely by microwave pulses, and this has been extensively experimentally demonstrated earlier\cite{Scrambling21,cervera2022experimental}. Here, we employ simple amplitude modulation of Gaussian square pulses for two qutrit operations, along with Gaussian with quadrature derivative pulses (DRAG) for single qutrit operations. This activates our two-qutrit gate, which is fully characterized and utilized to generate maximally entangled states. 
\par
\newcommand{\hdeff}{H_{d}^{\text{eff}}}
\par
\begin{figure}
\centering
\captionsetup{justification=RaggedRight,singlelinecheck=false}

\includegraphics[width=1\linewidth]{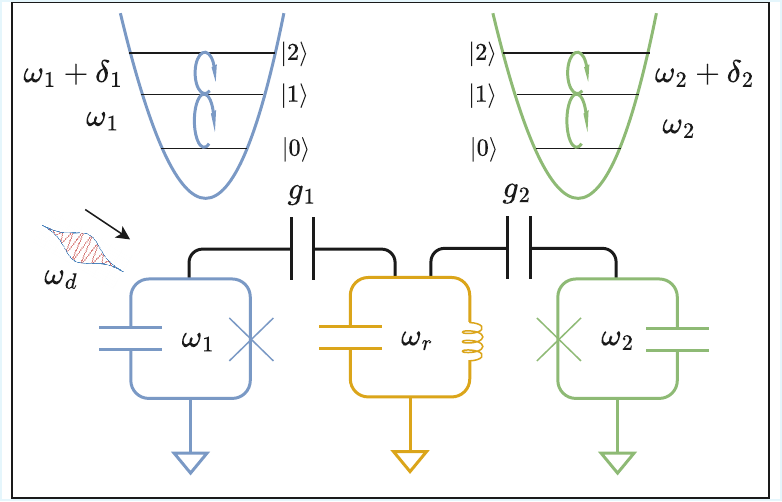}

\caption{The schematic diagram of a coupled two transmon system. $\omega_1$ and $\omega_2$ correspond to the frequency of the transition $\ket{0}\rightarrow \ket{1}$ in Transmon $1$ and Transmon $2$, respectively, and $\delta_1$ and $\delta_2$ are their respective anharmonicities. The two transmons are coupled through a resonator of frequency $\omega_r$ with coupling strength $J$. The proposed Generalized-Cross Resonance scheme generalizes the qubit-qubit Cross Resonance scheme to the qutrit setting, where the control qutrit (Transmon 1) is irradiated with a microwave pulse of frequency $\omega_d$ the equals either $\omega_2$ corresponding to $\ket{0}\rightarrow \ket{1}$ transition or, $\omega_2 + \delta_2$ corresponding to $\ket{1} \rightarrow \ket{2}$ transition of the target qutrit (Transmon 2).}
\label{fig:pop-plots}
\end{figure}

For superconducting architectures, one of the most prominent techniques to drive transmons is the cross-resonance (CR) technique \cite{Rigetti2010,Gambetta2011}. This technique has been explored and studied for the experimental realization of high-fidelity two-qubit quantum gates, for instance, the controlled-NOT gate \cite{Gambetta16, Gambetta17}. This amplitude-modulated microwave-only technique significantly reduces hardware complexity, as it allows control pulses to be delivered through the same lines used for sending standard single-qutrit control signals. 
Additionally, the cross-resonance gate offers a distinct approach to entangling operations by avoiding the need for frequency tuning or coupler modulation, typically required in other methods. Because it operates effectively with fixed-frequency transmon qutrits, the CR gate naturally avoids noise introduced by dynamic frequency adjustments. 
Transmon-based qutrits, known for their strong coherence properties (longer coherence times), low sensitivity to charge fluctuations, and high single-qutrit gate fidelities, are particularly compatible with this scheme. Hence, for large, fault-tolerant systems, the CR gate approach tends to be more robust and engineering-friendly. This has been the driving force for the cross-resonance gate to be widely adopted for implementing two-qutrit gates in superconducting architectures with fixed-frequency transmons \cite{Scrambling21}.

In this interaction, a microwave pulse is applied to one transmon (the control) at the resonant frequency of the other (the target). The interplay between the inherent coupling and the applied drive generates Rabi oscillations in the target qutrit, where the state of the control qutrit determines the oscillation frequency of the target qutrit. This conditional interaction forms the basis of the CR gate, enabling high-fidelity entangling operations without requiring direct frequency tuning.

\begin{figure*}[t]
\centering
\begin{subfigure}[b]{0.42\textwidth}
\fbox{
   \includegraphics[width=\linewidth]{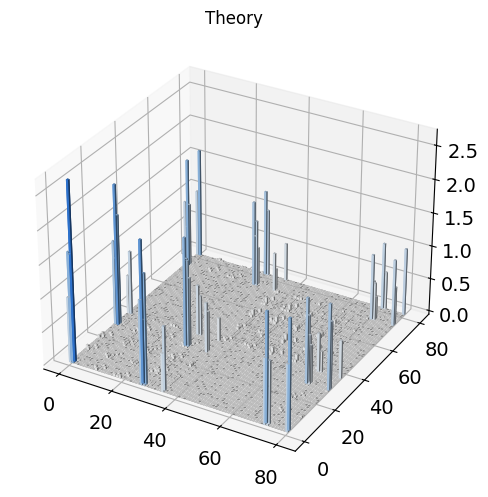}
   }
   \caption{}
\end{subfigure}
\hspace{0.5cm}
\begin{subfigure}[b]{0.42\textwidth}
    \fbox{
   \includegraphics[width=\linewidth]{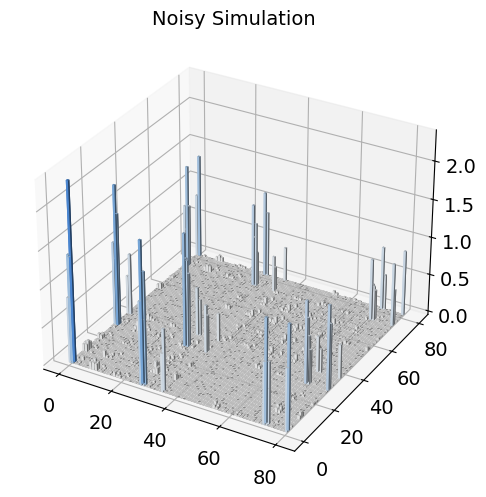}
   }
   \caption{}
\end{subfigure}

\begin{subfigure}[b]{0.42\textwidth}
\fbox{
   \includegraphics[width=\linewidth]{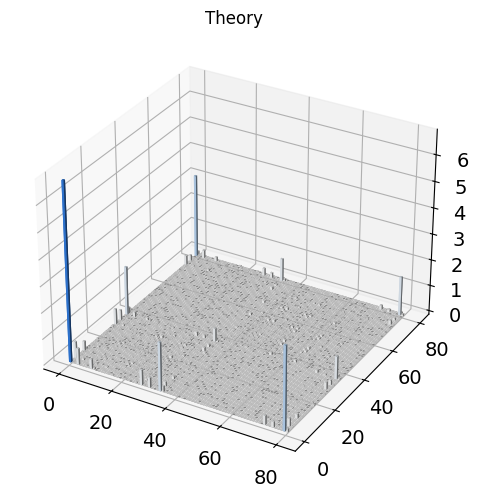}
   }
   \caption{}
\end{subfigure}
\hspace{0.5cm}
\begin{subfigure}[b]{0.42\textwidth}
    \fbox{
   \includegraphics[width=\linewidth]{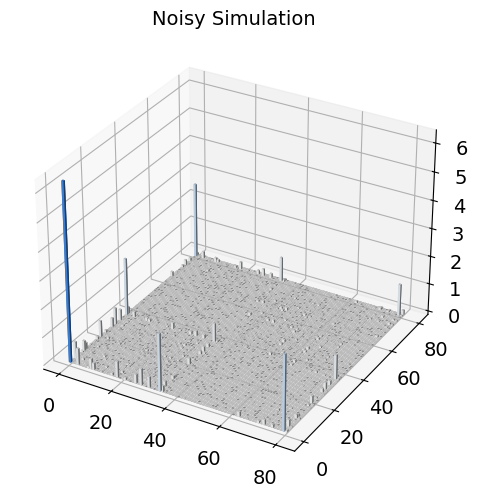}
   }
   \caption{}
\end{subfigure}

\captionsetup{justification=raggedright,singlelinecheck=false}

\caption{The comparison between the process matrices ($\chi_{ideal}$, $\chi_{noisy}$) of the ideal gates and the gates simulated under the effects of decoherence. (a) $\chi_{ideal}$ obtained using QPT for the $\gcxzero$ gate as defined in \ref{eqn:ucx01}. (b) $\chi_{noisy}$ obtained using QPT for the simulated $\gcxzero$ gate. We obtain a process fidelity of $86.14\% \pm 0.01\%$ for the simulated $\gcxzero$ gate when the decoherence time T ($T_1=T_2=T$) = $1700\mu s$. (c) $\chi_{ideal}$ obtained using QPT for the $\gcsqrtone$ gate as defined in \ref{eqn:uch12}. (d) $\chi_{noisy}$ obtained using QPT for the simulated $\gcsqrtone$ gate. We obtain a process fidelity of $91.99 \% \pm 0.01\%$ for the simulated $\gcsqrtone$ gate when the decoherence time T = $1700\mu s$.}

\label{fig:qpt-plots}
\end{figure*}


\begin{figure*}
    \centering
    \includegraphics[width=\linewidth]{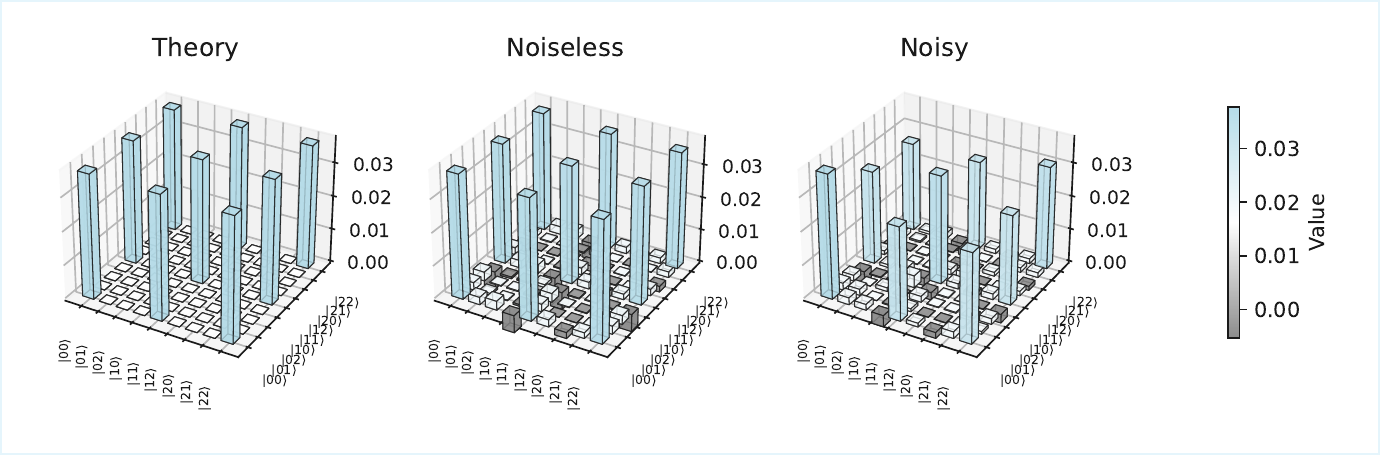}
    
    \captionsetup{justification=raggedright,singlelinecheck=false}
    \caption{ The QST plots of the (a) two-qutrit Bell state as defined in \ref{eqn:bell-state}, (b) the two-qutrit Bell state prepared under the noiseless setting, and (c) the two-qutrit Bell state prepared under the noisy setting ($T_1 = T_2 = 1700\mu s$). We prepare the two-qutrit Bell state using the proposed $\gcxzero$ and $\gcsqrtone$ gates and obtain state fidelities of $\mathcal{F}_{noiseless}=99.06 \pm 0.01\%$ and $\mathcal{F}_{noisy}=78.73 \pm 0.01\%$, repectively. For our choice of parameters, the total time for the preparation of the qutrit Bell state is $\sim514$ ns.}
    \label{fig:bell-plots}
\end{figure*}

For two-level systems, the cross-resonance technique, for a particular drive power, results in two distinct Rabi frequencies, $\gamma_0$ and $\gamma_1$. The Hamiltonian for this interaction can be described as
\begin{equation}
    \begin{aligned}
        H_{cr}^{\text{qubit}}=\gamma_0 |0\rangle \langle0| \otimes \sigma_{x} + \gamma_1 |1\rangle \langle1| \otimes \sigma_{x}, 
    \end{aligned}
    \label{eq1}
\end{equation}
where, $\gamma_0=w_{1}/2$ and $\gamma_1=-w_{1}/2$, with $w_{1}$ being the $\ket{0}\rightarrow \ket{1}$ transition frequency of qubit 1.

Extending the idea of cross-resonance to a two-qutrit system, we show, in the Supplementary Material \footnotemark[1]\footnotetext[1]{Refer to the Section I of the Supplementary Material.}, that the effective drive Hamiltonian of a microwave pulse on one qutrit with frequency resonant to the other qutrit can be given as, 
\begin{equation}
    \begin{aligned}
        &H_{cr}^{\text{qutrit}} (t) = \left(-\frac{J\Omega(t)}{4}\right) \bigg[ \bigg\{\nu_0 \ketbra{0} + \nu_1 \ketbra{1}+ \nu_2 \ketbra{2}\bigg\}  \\
        &\otimes \bigg\{ \eta_{20} \Big( e^{i(\omega_d - \omega_2^{01})t} \ketbra{0}{1} + e^{-i(\omega_d - \omega_2^{01})t} \ketbra{1}{0} \Big) \\
        &+ \sqrt{2}\eta_{21} \Big( e^{i(\omega_d - \omega_2^{12})t} \ketbra{1}{2} + e^{-i(\omega_d - \omega_2^{12})t} \ketbra{2}{1} \Big) \bigg\} \bigg]
    \end{aligned}
    \label{eq2}
\end{equation}

where, $\omega_d$ is the drive frequency, $\Omega (t)$ is the microwave-drive amplitude, along with, $\omega_2^{01} = \omega_2$, $\omega_2^{12} = \omega_2-\delta_2,$ $\nu_0 = -\eta_{10}, \nu_1 = \eta_{10}-2\eta_{11}$ and $\nu_2 = 2\eta_{11}$ with $\eta_{10} = 1/\omega_{1}$, $\eta_{11} = 1/(\omega_{1} + \delta_1)$, $\eta_{20}=1/\omega_{2}$, and $\eta_{21}=1/(\omega_{2} + \delta_2)$.      \\

It is apparent from the expression of $H_{cr}^{\text{qutrit}}(t)$ that the excitations in the target qutrit occur on the $0{-}1$ level if the drive frequency conforms to the $\ket{0}\rightarrow \ket{1}$ transition frequency of the target qutrit and occur on the $1{-}2$ level if the drive frequency matches the $\ket{1}\rightarrow \ket{2}$ transition frequency. 
Depending on the choice of the drive frequency, this Hamiltonian gives rise to one of the two gates, 
\begin{equation}
    \begin{aligned}
        \ucrzeroone(\theta) &= \sum_{i=0}^{2}\ketbra{i} \otimes R_X^{01}(\nu_i\theta)\\
        &=\ketbra{0}\otimes e^{-i(\ketbra{0}{1}+\ketbra{1}{0})\nu_0\theta}+\ketbra{1}\otimes \mathbb{I}\\
        &+\ketbra{2}\otimes e^{-i(\ketbra{0}{1}+\ketbra{1}{0})\nu_2\theta},
    \end{aligned}
    \label{eq3}
\end{equation}
or,
\begin{equation}
    \begin{aligned}
        \ucronetwo(\theta)&=\sum_{i=0}^{2} \ketbra{i} \otimes R_X^{12}(\nu_i\theta)\\
        &=\ketbra{0}\otimes e^{-i(\ketbra{1}{2}+\ketbra{2}{1})\nu_0\theta}+\ketbra{1}\otimes \mathbb{I}\\
        &+\ketbra{2}\otimes e^{-i(\ketbra{1}{2}+\ketbra{2}{1})\nu_2\theta}\text{.}\\
    \end{aligned}
    \label{eq4}
\end{equation}\\
Here the rotations are \!$R_X^{01}(\nu_i\theta)= e^{-i(\ketbra{0}{1} + \ketbra{1}{0})\nu_i\theta/2} \equiv RX^{01}(\phi_i)$ on $0{-}1$ space, and, on $1{-}2$ subspace, $R_X^{12}(\nu_i\theta) = e^{-i(\ketbra{1}{2} + \ketbra{2}{1})\nu_i\theta/2} \equiv RX^{12}(\phi_i)$.

These gates can be seen as direct analogues of the two-qubit CR gate on the $0{-}1$ and $1{-}2$ transition levels, respectively. In addition, they generalize the gates of the prior works \cite{Scrambling21,cervera2022experimental} that use the cross-resonance framework on just two of the three levels. 
Our work improves upon the prior works in two primary ways. First is the ability to perform a direct entanglement on the $1{-}2$ levels of the target qutrit. While the previous works use the ability to entangle on $0{-}1$ levels to perform entanglement on the $1{-}2$ levels, our gates bring in the ability to perform entanglement directly on the $1{-}2$ levels of the target qutrit.
Secondly, $\ucrzeroone$ and $\ucronetwo$ are not individual gates but rather a class of gates (i.e., they are parametric in nature) that directly allow entanglement on $0{-}1$ and $1{-}2$ levels and can be used to construct a multitude of entangling gates of interest. 
However, all gates proposed prior to this work are specific non-parametric gates. In addition to these contrasts, the action of $\ucrzeroone$ and $\ucronetwo$ on the target qutrit uniquely depends on the state of the control qutrit, and no two states of control induce the same oscillation on the target qutrit. This is evident both in the simulation results \footnotemark[2] and gate descriptions \ref{eq3} and \ref{eq4}. Although these gates individually act on a two-level subspace of the three-level systems, together they span the complete two-qutrit Hilbert space.

To demonstrate this maximal spanning, we use the generalized CR to construct controlled-$X$ and controlled-$H$ gates on the $0{-}1$ and $1{-}2$ subspaces, which we define as,
\begin{equation}
    \begin{aligned}
        U_{\text{CX}}^{01} =\ 
        &\ket{0}\bra{0} \otimes 
        \begin{bmatrix}
        0 & 1 & 0 \\
        1 & 0 & 0 \\
        0 & 0 & i
        \end{bmatrix}
        +
        \ket{1}\bra{1} \otimes 
        \begin{bmatrix}
        1 & 0 & 0 \\
        0 & 1 & 0 \\
        0 & 0 & 1
        \end{bmatrix} \\
        +
        &\ket{2}\bra{2} \otimes 
        \begin{bmatrix}
        0 & 1 & 0 \\
        1 & 0 & 0 \\
        0 & 0 & -i
        \end{bmatrix}
    \end{aligned}
    \label{eqn:ucx01}
\end{equation}
\begin{equation}
    \begin{aligned}
        U_{\text{CX}}^{12} =\ 
        &\ket{0}\bra{0} \otimes 
        \begin{bmatrix}
        i & 0 & 0 \\
        0 & 0 & 1 \\
        0 & 1 & 0
        \end{bmatrix}
        +
        \ket{1}\bra{1} \otimes 
        \begin{bmatrix}
        1 & 0 & 0 \\
        0 & 1 & 0 \\
        0 & 0 & 1
        \end{bmatrix} \\
        +
        &\ket{2}\bra{2} \otimes 
        \begin{bmatrix}
        -i & 0 & 0 \\
        0 & 0 & 1 \\
        0 & 1 & 0
        \end{bmatrix}
    \end{aligned}
    \label{eqn:ucx12}
\end{equation}
and
\begin{equation}
    \begin{aligned}
        U_{\text{CH}}^{01} =\ & \ket{0}\bra{0} \otimes
        \frac{1}{\sqrt{2}}\begin{bmatrix}
        1 & 1 & 0 \\
        -1 & 1 & 0 \\
        0 & 0 & \sqrt{2}
        \end{bmatrix}
        + \ket{1}\bra{1} \otimes
        \begin{bmatrix}
        1 & 0 & 0 \\
        0 & 1 & 0 \\
        0 & 0 & 1
        \end{bmatrix} \\
        & + \ket{2}\bra{2} \otimes
        \frac{1}{\sqrt{2}}\begin{bmatrix}
        1 & -1 & 0 \\
        1 & 1 & 0 \\
        0 & 0 & \sqrt{2}
        \end{bmatrix}
    \end{aligned}
    \label{eqn:uch01}
\end{equation}
\begin{equation}
    \begin{aligned}
        U_{\text{CH}}^{12} =\ & \ket{0}\bra{0} \otimes
        \frac{1}{\sqrt{2}}\begin{bmatrix}
        \sqrt{2} & 0 & 0 \\
        0 & 1 & 1 \\
        0 & -1 & 1
        \end{bmatrix}
        + \ket{1}\bra{1} \otimes
        \begin{bmatrix}
        1 & 0 & 0 \\
        0 & 1 & 0 \\
        0 & 0 & 1
        \end{bmatrix} \\
        & + \ket{2}\bra{2} \otimes
        \frac{1}{\sqrt{2}}\begin{bmatrix}
        \sqrt{2} & 0 & 0 \\
        0 & 1 & -1 \\
        0 & 1 & 1
        \end{bmatrix}
    \end{aligned}
    \label{eqn:uch12}
\end{equation}


We perform numerical simulations of the proposed gates using Qiskit Dynamics~\cite{qiskit_dynamics_2023}, both with and without the inclusion of decoherence effects. For effective decoherence time, we have considered Amplitude damping ($T_1$: Energy Relaxation Time) and Phase damping ($T_2$: Phase Coherence Time). We have also performed Quantum Process Tomography (QPT) and studied the fidelities of our proposed gates with respect to varying decoherence times \footnotemark[2]\footnotetext[2]{Refer to the Section V of the Supplementary Material.}. The noiseless simulations were performed by simulating the Hamiltonian dynamics, and the noisy simulations were performed by simulating the Lindbladian dynamics \footnotemark[3]\footnotetext[3]{The Lindblad operators corresponding to the qutrit amplitude and phase damping, which we have considered in this work, are as presented in the section IV of the supplementary material.}. 
For simplicity in simulations, we set the energy relaxation and the phase coherence times of both the $0-1$ levels and $1-2$ levels to be the same and across both qutrits.


Using QPT (Figure.\ref{fig:qpt-plots}), we obtain the process fidelities ($\mathcal{F}$) of these gates. Without decoherence effects, we obtain $\mathcal{F}({\gcxzero}) = 99.73 \pm 0.01\%$, $\mathcal{F}({\gcxone}) = 97.88 \pm 0.01\%$, $\mathcal{F}({\gcsqrtzero}) = 99.39 \pm 0.01\%$, and $\mathcal{F}({\gcsqrtone}) = 98.99 \pm 0.01\%$. Under the effects of decoherence, where $T_1 = T_2 = 1700\mu s$, we obtain fidelities of $\mathcal{F}({\gcxzero}) = 86.14 \pm 0.01\%$, $\mathcal{F}({\gcxone}) = 84.6 \pm 0.01\%$, $\mathcal{F}({\gcsqrtzero}) = 92.35 \pm 0.01\%$, and $\mathcal{F}({\gcsqrtone}) = 91.99 \pm 0.01\%$.
For comparison, we also obtain the process fidelities of identity gates of lengths $200$ ns and $100$ ns (gate times of $U_{CX}$ and $U_{CH}$, respectively) under the same noise parameters, and we obtain $\mathcal{F}(\mathbb{I}_{200}) = 86.87 \pm 0.01\%$ and $\mathcal{F}(\mathbb{I}_{100}) = 93.39 \pm 0.01\%$. Note that these fidelities are very close to the $U_{CX}$ and $U_{CH}$ gates, respectively.

Further, we use $\gcxzero$ and $\gcsqrtone$ gates, along with single qutrit gates, to demonstrate that the proposed generalized CR gates span the complete two-qutrit Hilbert space by preparing the two-qutrit Bell state:
\begin{equation}
    \ket{\psi} = \frac{1}{\sqrt{3}}(\ket{00} + \ket{11} + \ket{22})
    \label{eqn:bell-state}
\end{equation}
The exact pulse sequence of the Bell state preparation circuit can be found in the supplementary material \footnotemark[2].

The parameters used for the simulations mimic those of the actual backend, which were provided earlier. For the two-qutrit Bell state preparation, we use Quantum State Tomography (QST) to reconstruct the prepared states (Figure.\ref{fig:bell-plots}) and observe fidelities of $\mathcal{F}_{noiseless} = 99.06 \pm 0.01\%$ without decoherence and $\mathcal{F}_{noisy} = 78.73 \pm 0.01\%$ with decoherence, where $T_1 = T_2 = 1700\mu s$. For our simulator parameters, we obtain the total gate time corresponding to the Bell state preparation circuit as $\sim 514$ ns.

\par
In conclusion, we extend the qubit Cross-Resonance scheme to qutrit systems and show that it allows spanning the entire two-qutrit Hilbert space and achieving maximal entanglement.
To advocate this, we construct two-qutrit gates, namely $\gcxzero$, $\gcxone$, $\gcsqrtzero$ and $\gcsqrtone$, and use them with single qutrit gates to prepare a high-fidelity two-qutrit Bell state.
While previous works on the cross-resonance scheme for qutrits were limited to just two of the three dimensions of the qutrit Hilbert space~\cite{Scrambling21, cervera2022experimental}, our work improves upon them to show the possibility of exploring the complete qutrit Hilbert space using cross-resonance. Our technique of extending and generalizing the cross-resonance scheme also outperforms the modified cross-Kerr techniques of prior works in terms of gate times~\cite{Goss2022-nature, LuoExperimental}.

We further notice that, for our parameters of the simulation, we achieve a Bell state preparation time that is significantly lower than those due to prior works. This work throws light on one of the central concerns of gate timings that is posed on the CR scheme and shows the possibility of spanning much larger Hilbert spaces using considerably less gate times by generalizing the CR technique. This eliminates the need to invoke and work with more complex approaches for the current hardware architectures.
\par
A possible interesting direction of future work is to investigate if we can obtain a maximally-entangling gate on $d>2$ level systems using a single pulse. Further, while our work focuses on fixed-frequency transmon qutrits, recent advances in fluxonium qubit platforms \cite{PhysRevA.106.062602, PhysRevX.14.041014, PRXQuantum.5.040318} present interesting opportunities for future extension of GCR scheme. These investigations could be of paramount importance for the field of superconducting quantum computation and information processing on higher level systems, specifically qutrits. \\
\par

\noindent \textit{Acknowledgement:} Y.S. and R.K.L. are partially supported by the National Science and Technology Council of Taiwan (Nos 113-2123-M-007-003, 114-2112-M-007-044-MY3, 114-26229-007-005). Y.S., T.S. and S.C. would like to acknowledge the constant support that we received from Dr. Debajyoti Bera. S.C. and T.S. would also like to thank Harsh Rathee and Ming-Tso W. for extensive discussions on qutrits. Y.S. would also like to thank the Center for Quantum Technologies (CQT), IIIT-Delhi, for funding support in the initial phase of this work.
\\

\noindent \textit{Code Availability}: The code for all the simulations performed in this work are available at \href{https://github.com/braqiiit/Maximally-Entangling-Two-Qutrit-CR}{https://github.com/braqiiit/Maximally-Entangling-Two-Qutrit-CR}. \\

\bibliography{refs}
\newpage
\clearpage 
\onecolumngrid 

\appendix 

\section{Hamiltonian Calculation}

Generally, transmons are not considered ideal qubits as they exhibit lower anharmonicity and are best described as multi-level systems rather than simple two-level systems. Entangling gates can be implemented in superconducting architectures primarily using either tunable or fixed-frequency transmons\cite{engineer_guide}. In the case of tunable transmons, the coupling between the two transmons, referred to here as qutrits, can be controlled. This is achieved by tuning the qutrits in and out of resonance through flux tuning, allowing the coupling to be switched on or off. In contrast, fixed-frequency transmons rely on an ``always-on" interaction between the qutrits. This interaction is facilitated through microwave drives, and because of this external control, the transmons can be operated at ``
sweet spots" for extended periods. This increases the coherence times, making them compatible with current architectures.

To realize the generalized cross-resonance gates that we aim to implement as maximally entangling two-qutrit gates, we assume that the two qutrits are not in direct contact with one another; instead, they are individually coupled to a cavity. To study the effective interaction between the qutrits, we must apply the Schrieffer-Wolff transformation to derive the effective Hamiltonian, excluding the cavity terms.
The Hamiltonian for the two transmon system coupled via the resonator\cite{Thesis_RWTH},
\begin{equation}
    \begin{aligned}
    H &= \underbrace{\omega_r a^\dagger a + \sum_{i=1}^2 \sum_{j_i=0}^2 \left[ \left( \omega_i - \frac{\delta_i}{2} \right) j_i + \frac{\delta_i}{2} j_i^2 \right] \ket{j_i}\bra{j_i}}_{H_0} + \underbrace{\sum_{i=1}^2 g_i \left( a^\dagger c_i + a c_i^\dagger \right)}_{H_{\text{interaction}}} \\
    &= \omega_r a^\dagger a + \sum_{i=1}^2 \sum_{j_i=0}^1 \omega_{j_i} \ket{j_i}\bra{j_i} + \sum_{i=1}^2 g_i \left( a^\dagger c_i + a c_i^\dagger \right)
    \end{aligned}
    \label{eq:1}
\end{equation}

where, the index $i$ represents the transmon, while, $j_i$ is the energy level corresponding to $i^{th}$ transmon. Further, $\omega_r$ is the resonator frequency, $\omega_i$ are the respective transmon frequencies, $\delta_i$ are the respective anharmonicities and $g_i$ are the corresponding coupling constant between the resonator and the respective transmons. 

\begin{tikzpicture}
\draw [black, line width=0.1pt] (0,0) -- (17,0);
\end{tikzpicture}
\subsection*{Schrieffer-Wolff transformation}

The Schrieffer–Wolff (SW) transformation is a unitary transformation method used to derive an effective Hamiltonian $H_{\text{eff}}$ by decoupling low-energy and high-energy subspaces, enabling the construction of simplified effective models.

The starting point is a total Hamiltonian $H = H^0 + H'$, where $H^0$ denotes the unperturbed Hamiltonian with known eigenstates and eigenvalues, and $H'$ represents a small perturbation. The eigenstates of $H^0$ are assumed to split into two subsets (A and B) that are separated by a spectral gap $\Delta$. The perturbation $H'$ is considered sufficiently weak so as not to close this gap, satisfying a condition such as $||\epsilon V|| \leq \Delta / 2$ (where $H' = \epsilon V$). The perturbation term $H'$ is further decomposed into a block-diagonal component $H^1$ and a block-off-diagonal component $H^2$ with respect to the A and B subspaces.

The Schrieffer–Wolff transformation is expressed as
$$
H_{\text{eff}} = e^{-S} H e^S,  
$$
where $S$ is an anti-Hermitian operator selected such that the block-off-diagonal components between the A and B subspaces are eliminated in $H_{\text{eff}}$. By applying the Baker–Campbell–Hausdorff formula to expand $H_{\text{eff}}$, and requiring that the off-diagonal terms vanish at each order in the perturbative expansion $S = \sum_{j=1}^\infty S^{(j)}$, the operator $S$ can be systematically determined.
For instance, setting the first-order off-diagonal part of $H_{\text{eff}}$ to zero yields the equation relating $S^{(1)}$ (the first-order term in $S$) to the first-order block-off-diagonal part of $H'$, which is $H^2$:
$$
[H^0, S^{(1)}] = -H^2.
$$
Keeping terms up to second order in the perturbation $\epsilon$, the effective Hamiltonian is given by
$$
H_{\text{eff}} \approx H^0 + H^1 + \frac{1}{2} [H^2, S^{(1)}].
$$
This second-order term $\frac{1}{2} [H^2, S^{(1)}]$ effectively incorporates the influence of the high-energy states (mixed in by $H^2$) on the low-energy subspace. The operator $S$ is uniquely determined under certain conditions.

\begin{tikzpicture}
\draw [black, line width=0.1pt] (0,0) -- (17,0);
\end{tikzpicture}\\

As we see in the previous section, we need to choose a suitable ansatz corresponding to our Hamiltonian for the Schrieffer-Wolff (SW) transformation, 

\begin{equation}
    S^{(1)} = -\sum_{i=1}^2 g_i \left( a^\dagger \tilde{c}_i - a \tilde{c}_i^\dagger \right)
    \label{eq:2}
\end{equation}
where, 
\begin{equation}
    \tilde{c}_i = \sum_{j_i=0}^1 \frac{\sqrt{j_i+1}}{\omega_r - \omega_{j_i} - \delta_i j_i} \ket{j_i}\bra{j_i+1}
    \label{eq:3}
\end{equation}
The modified annihilation operator $\tilde{c}_i$ represents a dispersively dressed annihilation operator for qutrit $i$, when the transmon operates in the dispersive regime of transmon-resonator interactions. The modified annihilation operator $\tilde{c}_i$ accounts for the energy difference between the coupled states $\ket{n+1,j}$ and $\ket{n,j+1}$. It encapsulates effective transitions between adjacent qutrit levels mediated by a de-tuned resonator. After applying the transformation to equation \ref{eq:1}, we obtain the effective Hamiltonian,
\begin{equation}
    \begin{aligned}
        H_{\text{eff}} &= \left( \omega_r + \sum_{i=1}^2 \sum_{j_i=0}^2 \chi_{j_i} \ket{j_i}\bra{j_i} \right) a^\dagger a + \sum_{i=1}^2 \sum_{j_i=0}^1 \tilde{\omega}_{j_i} \ket{j_i}\bra{j_i} - \sum_{i \neq k} \frac{g_i g_k}{2} \left( c_i c_k^\dagger + \tilde{c}_i^\dagger \tilde{c}_k \right)
    \end{aligned}
    \label{eq:4}
\end{equation}
The quantity $\chi_{j_i} = g_{j_i}^2 (\mu_{j_i+1} - \mu_{j_i})$ represents the AC Stark shift, describing how the resonator's frequency is modified based on the transmon's energy level. Meanwhile, $\tilde{\omega}_{j_i} = \omega_{j_i} - g_{j_i}^2 \mu_{j_i}$ denotes the Lamb shift, which is the correction to the transmon’s own frequency due to its interaction with the cavity. Here, $\mu_{ji} = \frac{j_i}{\omega_r-\omega_i-\delta_i(j_i-1)}$. 

The coupling term, in equation \ref{eq:4}, can be re-expressed as shown below, where it describes transitions between joint transmon states. The coefficient $J_{j_1 j_2}$ captures the strength of this interaction and depends on the transmon frequencies, anharmonicities, and the resonator frequencies.
\begin{equation}
    \begin{aligned}
        H_{\text{coup}} &= - \sum_{i \neq k} \frac{g_i g_k}{2} (c_i \tilde{c}_k^\dagger + c_i^\dagger \tilde{c}_k) \\
        &= \sum_{j_1 j_2} \sqrt{j_1+1} \sqrt{j_2+1} J_{j_1 j_2} (\ket{j_1+1, j_2} \bra{j_1, j_2+1} + \ket{j_1, j_2+1} \bra{j_1+1, j_2}) \\ \quad \text{where,} \\
        J_{j_1 j_2} &= \frac{g_1 g_2 (\omega_1 + \omega_2 + \delta_1 j_1 + \delta_2 j_2 - 2\omega_r)}{2(\omega_r - \omega_1 - \delta_1 j_1) (\omega_r - \omega_2 - \delta_2 j_2)}
    \end{aligned}
    \label{eq:5}
\end{equation}
Under the approximation of low anharmonicity and low transmon number in the transmons, $J_{j_1 j_2}$ becomes approximately constant ($J_{j_1 j_2} \approx J$), and the modified annihilation operators $\tilde{c}_i$ reduce to the standard operators $c_i$.

As a result, the simplified coupling Hamiltonian becomes,
\begin{equation}
    H_{\text{coup}} = J [c_1^\dagger c_2 + c_1 c_2^\dagger]
    \label{6}
\end{equation}
and the full effective Hamiltonian includes this coupling along with terms accounting for the dispersive shifts in both the cavity and the transmon frequencies,

\begin{equation}
    \begin{aligned}
        H_{\text{eff}} &= \left[ w_r + \sum_{i=1}^2 \sum_{j_i=0}^2 \chi_j |j \rangle \langle j| \right] a^{\dagger} a + \sum_{i=1}^2 \sum_{j_i=0}^2 \tilde{w}_{ji} |j \rangle \langle j| + J \left[ c_1^{\dagger} c_2 + c_1 c_2^{\dagger} \right]
    \end{aligned}
    \label{7}
\end{equation}
At this stage, the cavity-related terms can be omitted since they only contribute an overall global phase and do not affect the system's dynamics.
\begin{equation}
    \begin{aligned}
        H_{\text{eff}} &= \sum_{i=1}^2 \sum_{j_i=0}^2 \tilde{w}_{ji} |j \rangle \langle j| + J \left[ c_1^{\dagger} c_2 + c_1 c_2^{\dagger} \right] \\
        &= \Big( \tilde{w}_{1,0} |0 \rangle \langle 0| + \tilde{w}_{1,1} |1 \rangle \langle 1| + \tilde{w}_{1,2} |2 \rangle \langle 2| \Big) \otimes \mathbb{I} + \mathbb{I} \otimes \Big( \tilde{w}_{2,0} |0 \rangle \langle 0| + \tilde{w}_{2,1} |1 \rangle \langle 1| + \tilde{w}_{2,2} |2 \rangle \langle 2| \Big) + J \left[ c_1^{\dagger} c_2 + c_1 c_2^{\dagger} \right]
    \end{aligned}
    \label{8}
\end{equation}
Restricting the system to the first three energy levels of each transmon, the Hamiltonian takes the form shown above, where each transmon's contribution is written as a tensor product with the identity operator on the other transmon, and the interaction is captured by the last term, $H_{\text{eff}}^{(2)}=J \left[ c_1^{\dagger} c_2 + c_1 c_2^{\dagger} \right]$.

When the qutrits are sufficiently detuned from the cavity, along with small coupling constant $J$, we can suitably apply a second Schrieffer-Wolff transformation using a suitable $S^{(1)}$ to approximately diagonalize the Hamiltonian. To proceed with this transformation on equation \ref{8}, it is necessary to evaluate the commutator $\left[ H_{\text{eff}}^{(2)}, S^{(1)} \right]$. A key step in this process is verifying that $\left[ H_{\text{eff}}^{(0)}, S^{(1)} \right] = -H^{(2)}$, which ensures the validity of the perturbative approach. \\

\begin{tikzpicture}
\draw [black, line width=0.1pt] (0,0) -- (17,0);
\end{tikzpicture}
\subsubsection*{\text{Showing} $\left[ H_{\text{eff}}^{(0)}, S^{(1)} \right] = -H^{(2)}$ \text{for SW-transformation}}
Here, we use the ansatz as, 
$$
S^{(1)} = -\frac{J}{2} (\hat{c}_1 \hat{c}_2^{\dagger} - \hat{c}_1^{\dagger} \hat{c}_2),
$$
where the annihilation and the creation operator are given by:
\begin{equation}
    \begin{aligned}
        \hat{c}_i = \eta_{i0} |0\rangle\langle 1| + \sqrt{2} \eta_{i1} |1\rangle\langle 2|, \quad \hat{c}_i^\dagger = \eta_{i0} |1\rangle\langle 0| + \sqrt{2} \eta_{i1} |2\rangle\langle 1|
    \end{aligned}
    \label{9}
\end{equation}
where, 
\begin{equation}
    \begin{aligned}
        \eta_{1j} = \frac{1}{\tilde{\omega}_{1,j+1}-\tilde{\omega}_{1,j}}, \quad \eta_{2j} = \frac{1}{\tilde{\omega}_{2,j+1}-\tilde{\omega}_{2,j}}
    \end{aligned}
    \label{10}
\end{equation}
For the unperturbed Hamiltonian, consider,
\begin{equation}
    \begin{aligned}
        &H_{\text{eff}}^{(0)} = \mathcal{A}_1 + \mathcal{B}_1, \ \text{where,}  &\mathcal{A}_1 = \tilde{\omega}_{11} |1\rangle\langle 1| + \tilde{\omega}_{12} |2\rangle\langle 2|, \ \mathcal{B}_1 = \tilde{\omega}_{21} |1\rangle\langle 1| + \tilde{\omega}_{22} |2\rangle\langle 2|
    \end{aligned}
    \label{11}
\end{equation}
We then compute the commutators, 
\begin{equation}
    \begin{aligned}
        -\frac{2[\mathcal{A}_1, S^{(1)}]}{J} = c_1 \hat{c}_2^\dagger + c_1^\dagger \hat{c}_2,
        \quad
        -\frac{2[\mathcal{B}_1, S^{(1)}]}{J} = -(\hat{c}_1^\dagger c_2 + \hat{c}_1 c_2^\dagger)
    \end{aligned}
    \label{12}
\end{equation}

Adding the two results, we get,  
\begin{equation}
    \begin{aligned}
        [H_{\text{eff}}^{(0)}, S^{(1)}] = -J (c_1 c_2^\dagger + c_1^\dagger c_2) = -H^{(2)},
    \end{aligned}
    \label{13}
\end{equation}
thus verifying the relation needed for the Schrieffer-Wolff transformation.

\begin{tikzpicture}
\draw [black, line width=0.1pt] (0,0) -- (17,0);
\end{tikzpicture}\\

Having verified the relation in Equation \ref{13}, we now compute the commutator $[H_{\text{eff}}^{(2)}, S^{(1)}]$, where the second-order Hamiltonian is given by

$$
H_{\text{eff}}^{(2)} = -J(c_1^\dagger c_2 + c_1 c_2^\dagger)
$$
We define the operators $\mathcal{A}_2 = c_1^\dagger c_2$ and $\mathcal{B}_2 = c_1 c_2^\dagger$. The commutator relation comes out to be,
\begin{equation}
    \begin{aligned}
        &[H_{\text{eff}}^{(2)}, S^{(1)}] \\
        &= -J([\mathcal{A}_2, S^{(1)}] + [\mathcal{B}_2, S^{(1)}]) \\
        &= -J^2 \Big(
        - \eta_{10} \eta_{20} \ket{01}\bra{01}
        - 2\eta_{10} \eta_{21} \ket{02}\bra{02} + \eta_{10} \eta_{20} \ket{10}\bra{10}
        + 2(\eta_{10} \eta_{21} - \eta_{11} \eta_{20}) \ket{11}\bra{11}\\
        &\quad- 4\eta_{11} \eta_{21} \ket{12}\bra{12}
        + 2\eta_{11} \eta_{20} \ket{20}\bra{20}
        + 4\eta_{11} \eta_{21} \ket{21}\bra{21} + (\eta_{11} \eta_{20} - \eta_{10} \eta_{21})(\ket{02}\bra{20} + \ket{20}\bra{02})
        \Big)
    \end{aligned}
    \label{14}
\end{equation}
To proceed with this analysis, we consider the effective static Hamiltonian, which is given by $H_0 + \frac{1}{2}[H_{(\text{eff})}^{(2)}, S^{(1)}]$, where $H_0$ is the unperturbed Hamiltonian. Evaluating the commutator leads to the following expression:
\begin{equation}
    \begin{aligned}
    \tilde{H}_{\text{eff}}
    &=H_0 + \frac{1}{2}[H_{\text{eff}}^{(2)}, S^{(1)}] \\
    &= \left(\tilde{\omega}_{21} + \left(\frac{J}{2}\right)^2 \eta_{10} \eta_{20} \right) \ket{01}\bra{01} + \left(\tilde{\omega}_{22} + J^2 \eta_{10} \eta_{21} \right) \ket{02}\bra{02} \\
    &\quad + \left(\tilde{\omega}_{11} - \frac{J^2}{2} \eta_{10} \eta_{20} \right) \ket{10}\bra{10} + (\tilde{\omega}_{11} + \tilde{\omega}_{22}) \ket{11}\bra{11} + \left(\tilde{\omega}_{11} + \tilde{\omega}_{22} + 2J^2 \eta_{11} \eta_{21} \right) \ket{12}\bra{12} \\
    &\quad + \left(\tilde{\omega}_{12} - J^2 \eta_{11} \eta_{20} \right) \ket{20}\bra{20} + \left(\tilde{\omega}_{12} + \tilde{\omega}_{21} - 2J^2 \eta_{11} \eta_{21} \right) \ket{21}\bra{21} + (\tilde{\omega}_{12} + \tilde{\omega}_{22}) \ket{22}\bra{22}
    \end{aligned}
\label{15}
\end{equation}
Here, each term corresponds to a computational basis state of the two-qutrit hilbert space, and the perturbative corrections shift the energy levels due to the effective qutrit-qutrit coupling. Under the assumption that the coupling is weak, i.e., $J^2 \ll \eta_{ij} \eta_{kl}$, the second-order corrections can be considered negligible. This allows us to approximate, $H_0 + [H^{(2)}_{\text{eff}}, S^{(1)}] \approx H_0$, greatly simplifying our analysis of the system in the rotating frame. In what follows, we will write down the explicit form of $H_0$ (originally presented in Equation \ref{eq:1}) and use it as the basis for incorporating the drive Hamiltonian and analyzing its effects.

We now revisit the foundational idea of the cross-resonance (CR) scheme, as mentioned in the main literature, where the target qutrit is indirectly controlled by applying a drive at a frequency that is resonant or nearly resonant with a particular transition of the control qutrit. The preceding analysis yields an effective Hamiltonian that sets the stage for implementing this scheme to realize our required two-qutrit entangling gate, the generalized cross-resonance gate. To proceed, it is necessary to introduce a microwave drive term—whose exact form will be addressed later—and to study its behavior in a more convenient frame of reference. This requires two essential transformations: first, we perform a Schrieffer–Wolff transformation to diagonalize the static effective Hamiltonian up to second order in the coupling strength $J$; second, we transition into a rotating frame at the frequency of the control qutrit. Constructing a suitable unitary operator $U(t)=e^{-iH_0t}$ for this step allows us to accurately describe the system's evolution in this rotating frame. Consider again, 
\begin{equation}
    \begin{aligned}
        H_0=\omega_r a^\dagger a + \sum_{i=1}^2 \sum_{j_i=0}^2 \left[ \left( \omega_i - \frac{\delta_i}{2} \right) j_i + \frac{\delta_i}{2} j_i^2 \right] \ket{j_i}\bra{j_i}
    \end{aligned}
    \label{16}
\end{equation}
In the context of the cross-resonance (CR) scheme, we must express the drive term in the reference frame of the control qutrit, which we denote by $i=2$. Since the overall dynamics of the system remain unaffected by the addition of a local phase, we can modify this Hamiltonian by introducing a unitary term. Restricting our attention to just the lowest three energy levels, this results in the modified form:
\begin{equation}
    \begin{aligned}
        \hat{H}_0&=\underset{i=2}{H_0} = w_2 \ket{1}\bra{1} + (2w_2 + \delta_2) \ket{2}\bra{2} - \frac{w_2}{2} \mathbb{I}\\
        &= -\left( \frac{w_2}{2} \ket{0}\bra{0} - \frac{w_2}{2} \ket{1}\bra{1} - \left(\frac{3w_2}{2} + \delta_2\right) \ket{2}\bra{2} \right)
    \end{aligned}
    \label{17}
\end{equation}
Using this, we can now write the suitable unitary for the change of reference frame, 
\begin{equation}
    \begin{aligned}
        U(t) &= e^{-i\hat{H}_0 t} \\
        &= e^{-i \left( \frac{\omega_2}{2} \right)t} |0 \rangle \langle 0| + e^{i \left( \frac{\omega_2}{2} \right)t} |1 \rangle \langle 1| + e^{i\left( \frac{3 \omega_2}{2} + \delta_2 \right) t} |2 \rangle \langle 2|   \\
        U^{\dagger}(t) &= e^{i\hat{H}_0 t} \\
        &= e^{i \left( \frac{\omega_2}{2} \right)t} |0 \rangle \langle 0| + e^{-i \left( \frac{\omega_2}{2} \right)t} |1 \rangle \langle 1| + e^{-i\left( \frac{3 \omega_2}{2} + \delta_2 \right) t} |2 \rangle \langle 2|
    \end{aligned}
    \label{18}
\end{equation}

Now, we construct a suitable microwave drive Hamiltonian, for driving the target qutrit, 
\begin{equation}
    \begin{aligned}
            H_d = \frac{\Omega(t)}{2} (c_1^\dagger e^{-i\hat{\omega}_2 t} + c_1 e^{i\hat{\omega}_2 t})
    \end{aligned}
    \label{19}
\end{equation}
The choice of the ansatz for applying Schrieffer–Wolff transformation to this Hamiltonian, is very similar to the one used before, $S^{(1)} = -\frac{J}{2} [\hat{c}_1 \hat{c}_2^\dagger - \hat{c}_1^\dagger\hat{c}_2]$. The diagonalized drive Hamiltonian after the transformation, let's say, $\tilde{H}_d$, is given by,
\begin{equation}
    \begin{aligned}
        &\tilde{H}_d=H_d+[H_d, S^{(1)}]\\
    \end{aligned}
    \label{20}
\end{equation}
where, 
\begin{equation}
    \begin{aligned}
        [H_d, S^{(1)}] &= -\left(\frac{J\Omega}{4}\right) 
        \Big[ (-\eta_{10} \ket{0}\bra{0} + (\eta_{10} - 2\eta_{11}) \ket{1}\bra{1} + 2\eta_{11} \ket{2}\bra{2}) \otimes (e^{-i\hat{\omega}_2 t} \hat{c}_2^\dagger + e^{i\hat{\omega}_2 t} \hat{c}_2) \Big]
    \end{aligned}
    \label{21}
\end{equation}
The next task is to transform this diagonalized drive hamiltonian to the target qutrit reference frame using the unitaries defined in equation \ref{18},
\begin{equation}
    \begin{aligned}
        \tilde{H}_d^{\text{eff}}=U(t)(H_d+[H_d, S^{(1)}])U^\dagger(t)
    \end{aligned}
    \label{22}
\end{equation}
Here, the gate corresponding to $U(t)H_dU^\dagger(t)$ acts as a single qutrit rotation just on the control qutrit, and hence has no contribution towards the entangling action. Further, it can be easily seen that $H_d \ \text{and }\ [H_d,S^{(1)}]$ commutates, hence, $e^{H_d+[H_d,S^{(1)}]}=e^{H_d}e^{[H_d,S^{(1)}]}$. Therefore, in order to get the required entangling two-qutrit gate, we rewrite, 
\begin{equation}
\begin{aligned}
\tilde{H}_d^{\text{eff}}(t) 
&= U(t)\left([H_d, S^{(1)}]\right)U^\dagger(t) \\
&= \left( -\frac{J\Omega(t)}{4} \right)\Bigg[ 
\left\{ -\eta_{10} \ket{0}\bra{0}
+ (\eta_{10} - 2\eta_{11}) \ket{1}\bra{1} \right. \left. + 2\eta_{11} \ket{2}\bra{2} \right\} \otimes 
\Bigg\{ \eta_{20} \left( e^{i(\hat{\omega}_2 - \omega_2)t} \ket{0}\bra{1} \right. \left. + e^{-i(\hat{\omega}_2 - \omega_2)t} \ket{1}\bra{0} \right) \\
&\quad  
+ \sqrt{2}\eta_{21} \left( e^{i(\hat{\omega}_2 - \omega_2 - \delta_2)t} \ket{1}\bra{2} \right. \left. + e^{-i(\hat{\omega}_2 - \omega_2 - \delta_2)t} \ket{2}\bra{1} \right) 
\Bigg\} \Bigg]\\
&= \left( -\frac{J\Omega(t)}{4} \right)\Bigg[ 
\Bigg\{ \nu_0 \ket{0}\bra{0}
+ \nu_1 \ket{1}\bra{1} + \nu_2 \ket{2}\bra{2} \Bigg\} \otimes 
\Bigg\{ \eta_{20} \left( e^{i(\omega_d - \omega_2^{01})t} \ket{0}\bra{1} \right. \left. + e^{-i(\omega_d - \omega_2^{01})t} \ket{1}\bra{0} \right) \\
&\quad  
+ \sqrt{2}\eta_{21} \left( e^{i(\omega_d - \omega_2^{12})t} \ket{1}\bra{2} \right. \left. + e^{-i(\omega_d - \omega_2^{12})t} \ket{2}\bra{1} \right) 
\Bigg\} \Bigg]  \\
&=H_{cr}^{\text{qutrit}}(t)
\end{aligned}
\label{23}
\end{equation}
where, $\omega_d=\hat{\omega}_2$ is the drive frequency, along with, $\omega_2^{01} = \omega_2$, $\omega_2^{12} = \omega_2-\delta_2,$ $\nu_0 = -\eta_{10}, \nu_1 = \eta_{10}-2\eta_{11}$ and $\nu_2 = 2\eta_{11}$ with $\eta_{10} = 1/\omega_{1}$, $\eta_{11} = 1/(\omega_{1} + \delta_1)$, $\eta_{20}=1/\omega_{2}$, and $\eta_{21}=1/(\omega_{2} + \delta_2)$. This hamiltonian can be used to obtain the desired gate by choosing appropriate values of $\omega_d (=\hat{\omega}_2) \ \text{as} \ \omega_2 \ \text{or} \ (\omega_2+\delta_2)$ as given in the main paper, 
\begin{equation}
    \begin{aligned}
        U_{\text{CR}}^{\text{01/12}}&=e^{-i\tilde{H}_d^{\text{eff}}t}=e^{-iH_{cr}^{\text{qutrit}}t}
    \end{aligned}
    \label{24}
\end{equation}

\subsection{List of approximations}

The derivation of the effective two-qutrit Hamiltonian in this section employs the following key approximations and assumptions. 

\begin{itemize}
    \item First, we operate in the dispersive regime, where both transmons are sufficiently detuned from the cavity mode such that direct energy exchange is suppressed. This condition is quantified by the requirement $|\Delta_i| \gg g_i$, where, $\Delta=\omega_r-\omega_i$ is the detuning between the cavity frequency $\omega_r$ and the transmon frequency $\omega_i$. This large detuning justifies the application of the Schrieffer–Wolff transformation to eliminate cavity degrees of freedom.
    \item Second, we assume a weak coupling regime where the perturbation parameter satisfies $\left\lVert H'\right\rVert \le \Delta / 2$, where $H'$ is represents a small perturbation to the initial Hamiltonian. This ensures the validity of the perturbative expansion in the Schrieffer–Wolff transformation. Specifically, we retain terms only up to second order: $H_{\text{eff}} \approx H^0 + H^1 + \frac{1}{2} [H^2, S^{(1)}]$, and systematically neglect higher-order corrections beyond this order. 
    \item Third, we operate in the low anharmonicity regime and perform a multi-level truncation to three levels per transmon. Each transmon is treated as a qutrit with anharmonicity parameters $\delta_i$ that are small enough to justify truncation, yet sufficiently nonzero to resolve three distinct energy levels. Under this assumption, the level-dependent coupling coefficients $J_{j_1j_2}$ (given in Eq. 5) become approximately constant: $J_{j_1j_2} \approx J$, and the dispersively-dressed operators $\tilde{c}_i$ reduce to standard creation/annihilation operators $c_i$.
    \item Fourth, we assume negligible direct qutrit–qutrit coupling. The two qutrits interact exclusively via cavity-mediated interactions; any direct coupling term in the full Hamiltonian is neglected. This is justified in fixed-frequency transmon architectures where qutrits are spatially separated and coupled only through the resonator. 
    \item Fifth, after the Schrieffer–Wolff transformation, we neglect cavity-induced global phases. The cavity operator terms that appear in the effective Hamiltonian (Eq. 7) contribute only an overall phase factor and do not affect the two-qutrit dynamics, permitting their removal.
    \item Sixth, to ensure the validity of the perturbative regime, we need to verify $[H^0_{\text{eff}},S^1]=-H^2$. For this, we assume the effective qutrit–qutrit coupling is weak and qutrits are sufficiently detuned from the cavity. This allows a second-order perturbative treatment with the ansatz $S^{(1)}=-\frac{J}{2}(\hat{c}_1\hat{c}_2^\dagger-\hat{c}_1^\dagger\hat{c}_2)$.
    \item Finally, we assume that second-order energy corrections are negligible. In deriving the final effective Hamiltonian (Eq. 15), the perturbative corrections arising from the commutator $\frac{1}{2}[H^{(2)}_{\text{eff}},S^{(1)}]$ are small compared to the system's energy scales. Specifically, we require $J^2 \ll \eta_{ij}\eta_{kl}$, where $\eta_{ij} = 1/(\tilde{\omega}_{i,j+1}-\tilde{\omega}_{i,j})$, permitting the approximation $H_0 + \frac{1}{2}[H^{(2)}_{\text{eff}}, S^{(1)}] \approx H_0$.
\end{itemize}

These collective approximations define the regime of validity and inherent limitations of the framework developed for implementing maximally entangling two-qutrit gates via the generalized cross-resonance scheme.\\

\section{Quantum Process Tomography}

To fully characterize an unknown quantum operation $\mathcal{E}$ acting on a finite-dimensional system, we use quantum process tomography. This involves preparing a complete set of $d^2$ linearly independent input states for a $d$-dimensional Hilbert space, applying $\mathcal{E}$ to each, and performing quantum state tomography \cite{Nielsen_Chuang_2010} on the corresponding outputs. The linearity of quantum operations allows reconstruction of $\mathcal{E}$'s full action from these input–output pairs. QPT provides a systematic method for experimentally determining the dynamics of quantum processes and is widely used to benchmark and validate quantum devices. The overall procedure is given by the Figure \ref{fig:qpt_flow}. 

\begin{figure}[ht]
\centering
\begin{tikzpicture}[node distance=0.9cm]

    \node (start) [mystep] {1. Choose system dimension \(d\) (e.g., \(d=3\) for a qutrit)};
    \node (stateprep) [mystep, below of=start] {2. Select \(d^2\) states \( \rho_0, \rho_1, \dots \rho_{d^2}\) that form a basis};
    \node (prep) [mystep, below of=stateprep] {3. Prepare the system in each state \(\rho_j\)};
    \node (process) [mystep, below of=prep] {4. Apply the unknown process \(\mathcal{E}\)};
    \node (tomography) [mystep, below of=process] {5. Perform quantum state tomography on \(\mathcal{E}(\rho_j)\)};
    \node (repeat) [mystep, below of=tomography] {6. Repeat steps 3–5 for all \(j\)};
    \node (reconstruct) [mystep, below of=repeat] {7. Reconstruct full process \(\mathcal{E}\) using tomography results};
    
    \draw [myarrow] (start) -- (stateprep);
    \draw [myarrow] (stateprep) -- (prep);
    \draw [myarrow] (prep) -- (process);
    \draw [myarrow] (process) -- (tomography);
    \draw [myarrow] (tomography) -- (repeat);
    \draw [myarrow] (repeat) -- (reconstruct);

\end{tikzpicture}
\caption{Step-by-step procedure for quantum process tomography.}
\label{fig:qpt_flow}
\end{figure}
To practically determine a useful representation of a quantum operation $\mathcal{E}$ from experimental data, we focus on the case of a single qubit. The objective is to identify a set of operation elements $\{E_i\}$ such that $\mathcal{E}$ can be expressed as:
\begin{equation}
   \mathcal{E}(\rho) = \sum_i E_i \rho E_i^\dagger 
\end{equation}
Since experimental outputs are numerical and not operator-based, it's more feasible to use a fixed set of operators $\{\tilde{E}_m\}$ forming a basis for operators on the state space. Each $E_i$ is then expressed in terms of this basis as:
\begin{equation}
    E_i = \sum_m e_{im} \tilde{E}_m,
\end{equation}
where $e_{im}$ are complex coefficients. Substituting this expansion into the earlier equation gives a new form of the operation:
\begin{equation}
    \mathcal{E}(\rho) = \sum_{mn} \tilde{E}_m \rho \tilde{E}_n^\dagger \chi_{mn},
\end{equation}
where $\chi_{mn} \equiv \sum_i e_{im} e_{in}^*$ and the matrix $\chi = [\chi_{mn}]$ is known as the chi matrix or the process matrix. This matrix is Hermitian and positive by definition. The process matrix representation enables complete characterization of the operation $\mathcal{E}$ in terms of a fixed operator basis, using only the matrix $\chi$ once the operator basis $\{\tilde{E}_m\}$ is chosen.
To determine the chi matrix $\chi$ for a quantum process, it’s important to recognize that although a general linear map on $d \times d$ matrices is defined by $d^4$ real parameters, physical constraints reduce the degree of freedom. Specifically, the density matrix $\rho$ must remain Hermitian and have trace one, introducing $d^2$ constraints. This is reflected in the completeness relation, $\sum_{i} E_i^\dagger E_i = I$. To find $\chi$ experimentally, we choose a basis $\{\rho_j\}$ of $d \times d$ matrices. 
Each output $\mathcal{E}(\rho_j)$ can be evaluated using the quantum state tomography\cite{Nielsen_Chuang_2010}, for each $\rho_j$ and further can be expressed as a linear combination of the basis elements:
\begin{equation}
    \mathcal{E}(\rho_j) = \sum_k \lambda_{jk} \rho_k,
\end{equation}
where $\lambda_{jk}$ are coefficients found using standard linear algebra methods. 
Before detailing the further steps of QPT, we explain quantum state tomography below.

\begin{tikzpicture}
\draw [black, line width=0.1pt] (0,0) -- (17,0);
\end{tikzpicture}
\subsection*{Quantum State Tomography}

The state $\rho_d$ of any $d$-dimensional quantum system can be represented in terms of a set of matrices that is orthonormal with respect to the Hilbert-Schmidt inner product.
For any such orthonormal set of matrices $\mathcal{B}_d = \{\hat{\sigma}_i : 0\le i\le d-1\}$, the state $\rho_d$ can be given as
$$\rho_d = \sum_{i=0}^{d-1} \lambda_i \hat{\sigma}_i.$$
Since the set $\{\lambda_i : 0\le i\le d-1\}$ of the coefficients uniquely describes the state $\rho_d$, the goal of quantum state tomography is to estimate the values of the coefficients accurately to get an accurate description of $\rho_d$.

For the case of qubits, the state of any single qubit $\rho_2$ can be represented in terms of the set of Pauli matrices $\mathcal{B}_2 = \{I, X, Y, Z\}$ as
\begin{equation}
    \begin{aligned}
        \rho_2&=\frac{I +\lambda_1X+\lambda_2Y+\lambda_3Z}{2}\\
        &=\frac{I+\text{tr}(X\rho)X+\text{tr}(Y\rho)Y+\text{tr}(Z\rho)Z}{2} 
    \end{aligned}
\end{equation}
For qutrits, one orthonormal set of matrices is the set of scaled Gell-Mann matrices $\mathcal{B}_3 = \{\hat{\sigma}_i : 0\le i\le 8\}$, where $\hat{\sigma}_0 = \frac{1}{\sqrt{3}}I$ and $\hat{\sigma}_i = \frac{1}{\sqrt{6}}\sigma_i$ with $\sigma_i$ being the $i^{th}$ Gell-Mann matrix.
Using this, the state $\rho_3$ can be represented as 

\begin{equation}
        \rho_3 =\sum_{i=0}^8 \lambda_i \hat{\sigma}_i    
        =\frac{1}{3}\Big(I + \sqrt{\frac{3}{2}}\sum_{i=1}^8 \text{tr}(\sigma_i\rho_3)\sigma_i\Big)
\end{equation}
Since the set $\{\text{tr}(\sigma_i\rho_3) : 1\le i\le 8\}$ of coefficients gives a complete description of the state $\rho_3$, quantum state tomography aims to estimate the values of the coefficients for any given state.
For two qutrit systems, as is in our case, the set of orthonormal set of matrices is the set $\mathcal{B}_9 = \mathcal{B}_3 \otimes \mathcal{B}_3 = \{\hat{\sigma}_i\otimes \hat{\sigma}_j : 0\le i, j \le 8\}$.
To get a complete description of the two-qutrit state, it is sufficient to estimate $\text{tr}\big((\hat{\sigma}_i\otimes \hat{\sigma}_j)\rho_9 \big)$ for all $0\le i, j\le 8$.

\begin{tikzpicture}
\draw [black, line width=0.1pt] (0,0) -- (17,0);
\end{tikzpicture}

Coming back to QPT, the action of $\tilde{E}_m$ and $\tilde{E}_n^\dagger$ on $\rho_j$ can be written as:

\begin{equation}
    \tilde{E}_m \rho_j \tilde{E}_n^\dagger = \sum_k \beta_{jk}^{mn} \rho_k,
\end{equation}
where $\beta_{jk}^{mn}$ are also determined from known operators. Substituting into the chi matrix formulation, we relate these to the measured $\lambda_{jk}$:
\begin{equation}
    \sum_k \sum_{mn} \chi_{mn} \beta_{jk}^{mn} \rho_k = \sum_k \lambda_{jk} \rho_k
\end{equation}
Because the $\rho_k$ form a basis, we equate coefficients directly:
\begin{equation}
    \sum_{mn} \beta_{jk}^{mn} \chi_{mn} = \lambda_{jk}
\end{equation}
This final relation provides the condition needed to solve for $\chi$. It shows how the process matrix $\chi$ is tied to measurable quantities, with $\chi$, $\lambda$, and $\beta$ forming a linear system that fully determines the quantum operation $\mathcal{E}$.

\section{Bell State Preparation}

To prepare a maximally entangled Bell state in the two-qutrit system, we begin with the initial state $\ket{00}$, where both the control and target qutrits are in their respective ground states. The first operation in the preparation sequence is the application of a qutrit Hadamard gate, denoted by $H_3$, on the control qutrit. This gate performs a uniform superposition over the qutrit basis states $\{\ket{0}, \ket{1}, \ket{2}\}$, and is defined as:

\begin{equation}
    H_3 = \frac{1}{\sqrt{3}} \begin{bmatrix}
    1 & 1 & 1 \\
    1 & \omega & \omega^2 \\
    1 & \omega^2 & \omega \\
    \end{bmatrix},
    \quad \text{where } \omega = e^{2\pi i/3}
\end{equation}

Applying $H_3$ on the control qutrit transforms the state as follows:
\begin{equation}
    \ket{00} \xrightarrow{H_3 \otimes I} \frac{1}{\sqrt{3}}(\ket{00} + \ket{10} + \ket{20})
\end{equation}

Following the initial superposition, the circuit proceeds by applying the first of two parametric entangling gates derived from the generalized cross-resonance scheme. By choosing a particular optimal parameter, we construct $U_{\text{CX}}^{01}$, which operates conditionally on the $0{-}1$ subspace of the qutrit Hilbert space and is defined explicitly as:

\begin{equation}
    \begin{aligned}
        U_{\text{CX}}^{01} =\ 
        &\ket{0}\bra{0} \otimes 
        \begin{bmatrix}
        0 & 1 & 0 \\
        1 & 0 & 0 \\
        0 & 0 & i
        \end{bmatrix}
        +
        \ket{1}\bra{1} \otimes 
        \begin{bmatrix}
        1 & 0 & 0 \\
        0 & 1 & 0 \\
        0 & 0 & 1
        \end{bmatrix} 
        +
        \ket{2}\bra{2} \otimes 
        \begin{bmatrix}
        0 & 1 & 0 \\
        1 & 0 & 0 \\
        0 & 0 & -i
        \end{bmatrix}
    \end{aligned}
\end{equation}

This operation acts non-trivially on the target qutrit depending on the state of the control qutrit. When applied to the intermediate superposition state, it effects the transformation:   
\begin{equation}
    \begin{aligned}
        \frac{1}{\sqrt{3}}(\ket{00}+ \ket{10} + \ket{20}) 
        \ \xrightarrow{U_{\text{CX}}^{01}} \ 
        \frac{1}{\sqrt{3}}(\ket{01} + \ket{10} + \ket{21})
    \end{aligned}
\end{equation}
The next operation involves applying a single-qutrit gate on the target qutrit. This gate, denoted by $V = H^{12}$, acts as a qubit Hadamard gate in the $\ket{1}{-}\ket{2}$ subspace on the target qutrit, while leaving the $\ket{0}$ level unaffected. Its matrix representation in the computational basis is given by:
\begin{equation}
    H^{12} = 
    \begin{bmatrix}
    1 & 0 & 0 \\
    0 & \frac{1}{\sqrt{2}} & \frac{1}{\sqrt{2}} \\
    0 & \frac{1}{\sqrt{2}} & -\frac{1}{\sqrt{2}} \\
    \end{bmatrix}
\end{equation}

Applying this gate to the second qutrit (i.e., the target), the two-qutrit state transforms as:

\begin{equation}
\begin{aligned}
    \frac{1}{\sqrt{3}}(\ket{01} + \ket{10} + \ket{21}) 
    \xrightarrow{I \otimes V} 
    \frac{1}{\sqrt{3}} \Bigg(
    \frac{1}{\sqrt{2}} \ket{0}(\ket{1} + \ket{2}) + \ket{1}\ket{0} + \frac{1}{\sqrt{2}} \ket{2}(\ket{1} + \ket{2}) 
    \Bigg)
\end{aligned}
\end{equation}
Next, we apply the second parametric two-qutrit gate, denoted as $U_{\text{CH}}^{12}$, which acts conditionally (on the $1{-}2$ subspace) on the control qutrit’s state and applies different unitary operations on the second qutrit accordingly. This gate is defined as:
\begin{equation}
    \begin{aligned}
        U_{\text{CH}}^{12} =\ & \ket{0}\bra{0} \otimes
        \frac{1}{\sqrt{2}}\begin{bmatrix}
        \sqrt{2} & 0 & 0 \\
        0 & 1 & 1 \\
        0 & -1 & 1
        \end{bmatrix}
        + \ket{1}\bra{1} \otimes
        \begin{bmatrix}
        1 & 0 & 0 \\
        0 & 1 & 0 \\
        0 & 0 & 1
        \end{bmatrix} + \ket{2}\bra{2} \otimes
        \frac{1}{\sqrt{2}}\begin{bmatrix}
        \sqrt{2} & 0 & 0 \\
        0 & 1 & -1 \\
        0 & 1 & 1
        \end{bmatrix}
    \end{aligned}
\end{equation}

Applying this gate to the state prepared in the previous step yields the transformed state:

\begin{equation}
    \begin{aligned}
        \frac{1}{\sqrt{3}}\Big(&\frac{1}{\sqrt{2}}\ket{0}(\ket{1} + \ket{2}) + \ket{1}\ket{0} + \frac{1}{\sqrt{2}}\ket{2}(\ket{1} +\ket{2})\Big)
        \xrightarrow{U_{\text{CH}}^{12}}
        \frac{1}{\sqrt{3}}(\ket{0}\ket{1} + \ket{1}\ket{0} +\ket{2}\ket{2})
    \end{aligned}
\end{equation}
Finally, to obtain the desired two-qutrit Bell state, we apply a single-qutrit Pauli-$X$ gate acting between levels $\ket{0}$ and $\ket{1}$, denoted as $X^{01}$, on the target (second) qutrit. This gate swaps the $\ket{0}$ and $\ket{1}$ levels, leaving $\ket{2}$ unchanged. Applying this gate results in:

\begin{equation}
    \begin{aligned}
        \frac{1}{\sqrt{3}}\Big(&\ket{0}\ket{1} + \ket{1}\ket{0} + \ket{2}\ket{2}\Big)
        \xrightarrow{I \otimes X^{01}}\frac{1}{\sqrt{3}}\left(\ket{0}\ket{0} + \ket{1}\ket{1} + \ket{2}\ket{2}\right)
    \end{aligned}
\end{equation}
Thus, we arrive at a maximally entangled two-qutrit Bell state.

\section{Lindblad Jump Operator Calculation}

We begin by defining the transition operators for the system, which describe the raising and lowering of states in a generalized Fock space:

{
\begin{equation}
    \sigma^+ = \sum_{n} \sqrt{n+1} |n+1\rangle \langle n| \quad \left| \quad \sigma^- = \sum_{n} \sqrt{n+1} |n\rangle \langle n+1| \right.
\end{equation}
}

\textbf{Qubit Phase Damping}      \\

The phase damping interaction can be modeled using an interaction Hamiltonian where the system operator $a^\dagger a$ (the number operator) couples with the environment (represented by $b$ operators):

\begin{equation}
    H = \chi a^\dagger a (b + b^\dagger)
\end{equation}

For a two-level system (qubit), we define the annihilation and creation operators as:

\begin{equation}
    a = |0\rangle \langle 1| ; \quad a^\dagger = |1\rangle \langle 0| ; \quad a^\dagger a = |1\rangle \langle 1|
\end{equation}

Assuming the environment transition $(b + b^\dagger)$ behaves like a Pauli-X operator ($X_b$), defined as $|0\rangle \langle 1| + |1\rangle \langle 0|$, we can write the total Hamiltonian in its tensor product form:

\begin{equation}
    H = \chi |1\rangle \langle 1| \otimes X_b = 
    \begin{bmatrix} 0 & 0 \\ 0 & 1 \end{bmatrix} \otimes 
    \begin{bmatrix} 0 & 1 \\ 1 & 0 \end{bmatrix}
\end{equation}

The resulting matrix representation of the interaction (ignoring the scalar $\chi$ for simplicity in the visualization) shows a block-diagonal structure where the interaction only affects the $|1\rangle$ subspace: 

\begin{equation}
    e^{-i (|1\rangle \langle 1| \otimes X_b) t} = 
    \begin{bmatrix} 
        0 & 0 \\ 
        0 & \begin{pmatrix} 0 & 1 \\ 1 & 0 \end{pmatrix} 
    \end{bmatrix}
\end{equation}

Next, we define the unitary evolution operator $U$ on the system-environment space. This operator applies a rotation to the system when the environment is in the $|1\pm\rangle$ states and acts as an identity when the environment is in the $|0\rangle$ subspace:

\begin{equation}
    U = e^{it} |1-\rangle \langle 1-| + e^{-it} |1+\rangle \langle 1+| + |01\rangle \langle 01| + |00\rangle \langle 00|
\end{equation}

The Kraus operators $E_k$ describe the effect on the system after tracing out the environment, assuming the environment starts in the $|0\rangle$ state:

\begin{equation}
    E_k = \langle e_k | U | 0 \rangle \quad \left| \quad \frac{e^{it} + e^{-it}}{2} = \cos t \right.
\end{equation}

Using the above identity, we calculate the \textbf{first Kraus operator $E_0$}:

\begin{equation}
\begin{aligned}
E_0 &= \langle 0 | U | 0 \rangle = \frac{e^{it}}{2} |1\rangle \langle 1| + \frac{e^{-it}}{2} |1\rangle \langle 1| + |0\rangle \langle 0| = |0\rangle \langle 0| + \cos t |1\rangle \langle 1|
\end{aligned}
\end{equation}

Similarly, we find the \textbf{second Kraus operator $E_1$}:

\begin{equation}
\begin{aligned}
E_1 &= \langle 1 | U | 0 \rangle = \frac{-e^{it}}{2} |1\rangle \langle 1| + \frac{e^{-it}}{2} |1\rangle \langle 1| = -\sin t |1\rangle \langle 1|
\end{aligned}
\end{equation}

In the computational basis $\{|0\rangle, |1\rangle\}$, these operators can be expressed in matrix form:

\begin{equation}
    E_0 = \begin{bmatrix} 1 & 0 \\ 0 & \cos t \end{bmatrix} ; \quad 
    E_1 = \begin{bmatrix} 0 & 0 \\ 0 & -\sin t \end{bmatrix}
\end{equation}

Finally, we can apply a unitary transformation $u_{ij}$ (specifically using the Hadamard basis) to find a new set of Kraus operators $F_i$:

\begin{equation}
    F_i = \sum_{j} u_{ij} E_j ; \quad 
    \frac{1}{\sqrt{2}} \begin{bmatrix} 1 & 1 \\ 1 & -1 \end{bmatrix} 
    \begin{pmatrix} E_0 \\ E_1 \end{pmatrix}
\end{equation}

By substituting our matrix forms, we derive $F_0$ and $F_1$:

\begin{align}
    F_0 &= \frac{1}{\sqrt{2}} \left( \begin{bmatrix} 1 & 0 \\ 0 & \cos t \end{bmatrix} + \begin{bmatrix} 0 & 0 \\ 0 & -i \sin t \end{bmatrix} \right) = \frac{1}{\sqrt{2}} \begin{bmatrix} 1 & 0 \\ 0 & e^{-it} \end{bmatrix} \\
    F_1 &= \frac{1}{\sqrt{2}} \left( \begin{bmatrix} 1 & 0 \\ 0 & \cos t \end{bmatrix} - \begin{bmatrix} 0 & 0 \\ 0 & -i \sin t \end{bmatrix} \right) = \frac{1}{\sqrt{2}} \begin{bmatrix} 1 & 0 \\ 0 & e^{it} \end{bmatrix}
\end{align}

This transformation corresponds to a change in the measurement basis of the environment, where the new operators relate to the rotated basis $|\tilde{b}_j\rangle$:

\begin{equation}
    F_{ik} = \sum_j U_{ij} E_{jk} = \left( \sum_j U_{ij} \langle b_j | U \right) |0\rangle = \langle \tilde{b}_j | U | 0 \rangle
\end{equation}

the state of the system after the unitary evolution can be expressed as:

\begin{equation}
    U |0\rangle = \frac{e^{it}}{\sqrt{2}} |1\rangle \langle 1| \otimes |-\rangle + \frac{e^{-it}}{\sqrt{2}} |1\rangle \langle 1| \otimes |+\rangle + |0\rangle \langle 0| \otimes |0\rangle
\end{equation}

To understand the relationship between the system's evolution and the identity, we express a scaled identity matrix as a combination of the Kraus operators $E_0$ and $E_1$:

\begin{equation}
    \sqrt{\frac{1+\sqrt{1-\lambda}}{2}} \begin{bmatrix} 1 & 0 \\ 0 & 1 \end{bmatrix} = 
    a \begin{bmatrix} 1 & 0 \\ 0 & \sqrt{1-\lambda} \end{bmatrix} + 
    b \begin{bmatrix} 0 & 0 \\ 0 & \sqrt{\lambda} \end{bmatrix}
\end{equation}

We can determine the relationship between $b$ and $a$:

\begin{equation}
    a = a\sqrt{1-\lambda} + b\sqrt{\lambda} \Rightarrow b = \frac{1-\sqrt{1-\lambda}}{\sqrt{\lambda}} a
\end{equation}

where, $a\sqrt{\frac{1+\sqrt{1-\lambda}}{2}}$   \\

\textbf{Qubit Amplitude Damping}      \\

The interaction between the system and the environment is modeled by the following interaction Hamiltonian, where $a$ and $b$ are the annihilation operators for the system and environment, respectively:

\begin{equation}
    H = \chi (a^\dagger b + b^\dagger a)
\end{equation}

For a 3-level system, the relaxation processes (decays) between different energy levels are captured by a set of Kraus operators. These operators describe the probabilistic evolution of the system state. The first operator, $K_0$, represents the case where no decay occurs, maintaining the population in the ground state while scaling the amplitudes of the excited states based on their decay rates:

\begin{equation}
    K_0 = \begin{bmatrix} 
        1 & 0 & 0 \\ 
        0 & \sqrt{1 - \gamma_{10}} & 0 \\ 
        0 & 0 & \sqrt{1 - \gamma_{21} - \gamma_{20}} 
    \end{bmatrix}
\end{equation}

The operator $K_1$ represents the single-step relaxation transitions, specifically the transitions from the first excited state to the ground state ($|1\rangle \rightarrow |0\rangle$) and the second excited state to the first ($|2\rangle \rightarrow |1\rangle$):

\begin{equation}
    K_1 = \begin{bmatrix} 
        0 & \sqrt{\gamma_{10}} & 0 \\ 
        0 & 0 & \sqrt{\gamma_{21}} \\ 
        0 & 0 & 0 
    \end{bmatrix}
\end{equation}

The operator $K_2$ accounts for the direct relaxation from the second excited state to the ground state ($|2\rangle \rightarrow |0\rangle$):

\begin{equation}
    K_2 = \begin{bmatrix} 
        0 & 0 & \sqrt{\gamma_{20}} \\ 
        0 & 0 & 0 \\ 
        0 & 0 & 0 
    \end{bmatrix}
\end{equation}

\textbf{The Lindblad Equation} \\

The time evolution of the system's density matrix $\rho$ is governed by the following equation, which combines unitary evolution with a dissipative term:

\begin{equation}
    \frac{\partial \rho}{\partial t} = -\frac{i}{\hbar} [H, \rho] + D[\rho]
\end{equation}

In this expression, the commutator $[H, \rho]$ represents the coherent, unitary part of the dynamics governed by the Hamiltonian $H$, while $D[\rho]$ represents the dissipator, accounting for noise and decoherence. For a system of dimension $d$, the dissipative part of the equation must satisfy a specific mathematical structure to ensure the density matrix remains physical (positive and trace-one):

\begin{equation}
    \frac{\partial \rho}{\partial t} = \sum_{k=1}^{d-1} L_k \rho L_k^\dagger - \frac{1}{2} \{L_k^\dagger L_k, \rho\}
\end{equation}

where $L_k$ are the jump operators.     \\

\textbf{Qutrit amplitude damping}    \\

For a 3-level system undergoing amplitude damping, we define a set of Kraus operators that describe the transitions between the ground state $|0\rangle$, the first excited state $|1\rangle$, and the second excited state $|2\rangle$:

\begin{equation}
\begin{aligned}
K_0 &=
\begin{bmatrix}
1 & 0 & 0 \\
0 & \sqrt{1-\gamma_{10}} & 0 \\
0 & 0 & \sqrt{1-\gamma_{20}-\gamma_{21}}
\end{bmatrix}
\; ; \;
K_1 =
\begin{bmatrix}
0 & \sqrt{\gamma_{10}} & 0 \\
0 & 0 & 0 \\
0 & 0 & 0
\end{bmatrix}
\; ; \;
K_2 =
\begin{bmatrix}
0 & 0 & \sqrt{\gamma_{20}} \\
0 & 0 & 0 \\
0 & 0 & 0
\end{bmatrix}
\; ; \;
K_3 =
\begin{bmatrix}
0 & 0 & 0 \\
0 & 0 & \sqrt{\gamma_{21}} \\
0 & 0 & 0
\end{bmatrix}
\end{aligned}
\end{equation}



The action of the noise over an infinitesimal time interval $dt$ transforms the density matrix $\rho$ as follows:

\begin{equation}
    \rho \longmapsto \sum_{i} K_i \rho K_i^\dagger = \rho(t + dt)
\end{equation}

To relate these to the Lindblad form, we assume the decay probabilities are proportional to the time interval, $\gamma_a = \gamma'_a dt$. Using a first-order Taylor expansion for the diagonal elements of $K_0$, we have $\gamma_a = \gamma'_a dt$ and $\sqrt{1 - \gamma_a} \approx 1 - \frac{\gamma_a}{2}$

Substituting the time-dependent rates into the transition operators allows us to express them in terms of the system's basis states:

\begin{equation}
\begin{aligned}
K_1 &= \sqrt{\gamma'_{10} dt} |0\rangle \langle 1| \; ; \;
K_2 = \sqrt{\gamma'_{20} dt} |0\rangle \langle 2| \; ; \;
K_3 = \sqrt{\gamma'_{21} dt} |1\rangle \langle 2|
\end{aligned}
\end{equation}

By factoring out $\sqrt{dt}$, we identify the Lindblad Jump Operators $L_i$, which represent the rate of specific decay processes:

\begin{equation}
\begin{aligned}
L_1 &= \sqrt{\gamma'_{10}} |0\rangle \langle 1| \Rightarrow K_1 = L_1 \sqrt{dt} \; ; \;
L_2 = \sqrt{\gamma'_{20}} |0\rangle \langle 2| \Rightarrow K_2 = L_2 \sqrt{dt} \; ; \;
L_3 = \sqrt{\gamma'_{21}} |1\rangle \langle 2| \Rightarrow K_3 = L_3 \sqrt{dt}
\end{aligned}
\end{equation}

Next, we express the Kraus operator $K_0$ in the computational basis $\{|0\rangle, |1\rangle, |2\rangle\}$. Using the infinitesimal approximation $\sqrt{1-\gamma_a dt} \approx 1 - \frac{\gamma'_a dt}{2}$, we expand the diagonal terms:

\begin{equation}
\begin{aligned}
K_0 &= |0\rangle \langle 0| + \sqrt{1-\gamma_{10}} |1\rangle \langle 1| + \sqrt{1-(\gamma_{20} + \gamma_{21})} |2\rangle \langle 2| \\
&= |0\rangle \langle 0| + \left(1 - \frac{\gamma'_{10} dt}{2}\right) |1\rangle \langle 1| + \left(1 - \frac{(\gamma'_{20} + \gamma'_{21}) dt}{2}\right) |2\rangle \langle 2|
\end{aligned}
\end{equation}

To simplify the subsequent algebra, we define auxiliary operators $A_i$ that capture the first-order decay effects on the excited state populations:

\begin{equation}
    A_1 = -\frac{\gamma'_{10}}{2} |1\rangle \langle 1|; \quad
    A_2 = -\frac{\gamma'_{20}}{2} |2\rangle \langle 2| ; \quad A_3 = -\frac{\gamma'_{21}}{2} |2\rangle \langle 2|
\end{equation}

This allows us to write $K_0$ as an identity operator perturbed by a small dissipative term proportional to $dt$:

\begin{equation}
    K_0 = \mathbb{I} + (A_1 + A_2 + A_3) dt
\end{equation}

The evolution of the density matrix over a step $dt$ is given by $\rho(t + dt) = \sum_i K_i \rho K_i^\dagger$. Substituting our expression for $K_0$ and the previously defined jump operators $K_i = L_i \sqrt{dt}$ for $i > 0$, we expand to first order in $dt$:

\begin{equation}
\begin{aligned}
\rho(t+dt) &= \sum_{i} K_i \rho K_i^\dagger \\
&\approx [\mathbb{I} + (A_1+A_2+A_3)dt] \rho [\mathbb{I} + (A_1^\dagger+A_2^\dagger+A_3^\dagger)dt] + L_1 \rho L_1^\dagger dt + L_2 \rho L_2^\dagger dt + L_3 \rho L_3^\dagger dt \\
&\approx \rho + (A_1\rho + A_2\rho + A_3\rho) dt + (\rho A_1^\dagger + \rho A_2^\dagger + \rho A_3^\dagger) dt + L_1 \rho L_1^\dagger dt + L_2 \rho L_2^\dagger dt + L_3 \rho L_3^\dagger dt
\end{aligned}
\end{equation}

Finally, by rearranging the equation into a finite difference form, we arrive at the differential generator of the noise process:

\begin{equation}
    \frac{\rho(t+dt) - \rho(t)}{dt} \approx \sum_{i=1}^{3} L_i \rho L_i^\dagger + A_i \rho + \rho A_i^\dagger
\end{equation}

By comparing the auxiliary operators $A_i$ with the definition of the jump operators $L_i$, we can express the dissipative terms as products of the jump operators and their adjoints:

\begin{equation}
\begin{aligned}
A_1 &= -\frac{\gamma'_{10}}{2} |1\rangle \langle 1| = -\frac{L_1^\dagger L_1}{2} \; ; \;
A_2 = -\frac{\gamma'_{20}}{2} |2\rangle \langle 2| = -\frac{L_2^\dagger L_2}{2} \; ; \;
A_3 = -\frac{\gamma'_{21}}{2} |2\rangle \langle 2| = -\frac{L_3^\dagger L_3}{2}
\end{aligned}
\end{equation}

Substituting these expressions into the master equation derived in the previous section, the terms $\sum (A_i \rho + \rho A_i^\dagger)$ naturally form an anti-commutator:

\begin{equation}
\begin{aligned}
\sum_i (A_i \rho + \rho A_i^\dagger) &= \sum_i \left( -\frac{L_i^\dagger L_i}{2} \rho - \rho \frac{L_i^\dagger L_i}{2} \right) = -\frac{1}{2} \sum_i \{L_i^\dagger L_i, \rho\}
\end{aligned}
\end{equation}

This yields the complete differential master equation describing the noise evolution of the qutrit:

\begin{equation}
    \frac{\rho(t+dt) - \rho(t)}{dt} \approx \sum_{i=1}^{3} L_i \rho L_i^\dagger - \frac{1}{2} \{L_i^\dagger L_i, \rho\}
\end{equation}

\textbf{Qutrit Phase damping}       \\

For this noise model, the Kraus operators $K_i$ are defined using combinations of the identity operator and state-specific phase shifts:

\begin{equation}
\begin{aligned}
K_0 &= \sqrt{1 - \gamma_1 - \gamma_2} \, \mathbb{I} \; ; \;
K_1 = \sqrt{\gamma_1} \left( |0\rangle \langle 0| - |1\rangle \langle 1| + |2\rangle \langle 2| \right) \; ; \;
K_2 = \sqrt{\gamma_2} \left( |0\rangle \langle 0| + |1\rangle \langle 1| - |2\rangle \langle 2| \right)
\end{aligned}
\end{equation}

We verify the completeness relation ($\sum K_i^\dagger K_i = \mathbb{I}$) to ensure the evolution is trace-preserving:

\begin{equation}
\begin{aligned}
K_0 K_0^\dagger + K_1 K_1^\dagger + K_2 K_2^\dagger &= (1 - \gamma_1 - \gamma_2) \mathbb{I} + \gamma_1 \mathbb{I} + \gamma_2 \mathbb{I} = \mathbb{I}
\end{aligned}
\end{equation}

To transition to a master equation representation, we assume the decay parameters are proportional to an infinitesimal time step $dt$, such that $\gamma_i = \gamma_i' dt$. We define the Lindblad Jump Operators $L_1$ and $L_2$ as:

\begin{equation}
\begin{aligned}
L_1 &= \sqrt{\gamma_1'} \left( |0\rangle \langle 0| - |1\rangle \langle 1| + |2\rangle \langle 2| \right) \; ; \;
L_2 = \sqrt{\gamma_2'} \left( |0\rangle \langle 0| + |1\rangle \langle 1| - |2\rangle \langle 2| \right)
\end{aligned}
\end{equation}

Using the first-order Taylor expansion $\sqrt{1-x} \approx 1 - \frac{x}{2}$, the $K_0$ operator can be approximated for small $dt$:

\begin{equation}
\begin{aligned}
K_0 &= \sqrt{1 - \gamma_1 - \gamma_2} \, \mathbb{I} \approx \left( 1 - \frac{\gamma_1 + \gamma_2}{2} \right) \mathbb{I} = \left( 1 - \frac{(\gamma_1' + \gamma_2')dt}{2} \right) \mathbb{I}
\end{aligned}
\end{equation}

Additionally, these jump operators satisfy the following property proportional to the identity:

\begin{equation}
    L_1^\dagger L_1 = \gamma_1' \mathbb{I} \quad \text{and} \quad L_2^\dagger L_2 = \gamma_2' \mathbb{I}
\end{equation}

To derive the continuous-time evolution, we first define auxiliary operators $A_1$ and $A_2$ that represent the first-order corrections to the identity operator in the Kraus representation:

\begin{equation}
    A_1 = -\frac{\gamma_1'}{2} \mathbb{I} \quad \text{and} \quad A_2 = -\frac{\gamma_2'}{2} \mathbb{I}
\end{equation}

Using these definitions, the non-jump Kraus operator $K_0$ is approximated as a perturbed identity:$$K_0 \approx \mathbb{I} + (A_1 + A_2) dt.$$ The evolution of the density matrix over an infinitesimal step $dt$ is given by $\rho(t + dt) = \sum K_i \rho K_i^\dagger$. Substituting our first-order approximations for $K_0$ and the jump operators $L_i$ (where $K_i = L_i \sqrt{dt}$), we obtain:

\begin{equation}
\begin{aligned}
\rho(t+dt) &= \sum K_i \rho K_i^\dagger \\
&\approx [\mathbb{I} + (A_1 + A_2) dt] \rho [\mathbb{I} + (A_1^\dagger + A_2^\dagger) dt] + L_1 \rho L_1^\dagger dt + L_2 \rho L_2^\dagger dt \\
&\approx \rho + (A_1 \rho + A_2 \rho) dt + (\rho A_1^\dagger + \rho A_2^\dagger) dt + \sum_i L_i \rho L_i^\dagger dt
\end{aligned}
\end{equation}

By rearranging this into a differential quotient, we see the generator of the dynamics:

\begin{equation}
    \frac{\rho(t+dt) - \rho(t)}{dt} = \sum_i L_i \rho L_i^\dagger + A_i \rho + \rho A_i^\dagger
\end{equation}

Finally, observing that $L_i^\dagger L_i = \gamma_i' \mathbb{I} = -2 A_i$, we substitute back to reach the final Lindblad Master Equation. Using $L_i^\dagger L_i = \gamma_i' \mathbb{I} = -2 A_i$, we get:

\begin{equation}
    \frac{\rho(t+dt) - \rho(t)}{dt} = \sum_i L_i \rho L_i^\dagger - \frac{1}{2} \{L_i^\dagger L_i, \rho\}
\end{equation}

\section{Simulations}

\subsection{Simulation Setup}

\subsubsection*{Qutrits parameters}

\begin{table}[H]
\centering
\begin{tabular}{|l|c|c|}
\hline
\textbf{Parameters (GHz)} & \textbf{$Q_1$} & \textbf{$Q_2$} \\
\hline
Natural frequency $\omega/2\pi$ & 4.9 & 5.5 \\
Anharmonicity $\alpha/2\pi$ & -0.4 & -0.3 \\
Individual coupling strength $g_{i}/2\pi$ & 0.62 & 0.64 \\

Qutrit-Qutrit direct coupling $J/2\pi$ & \multicolumn{2}{c|}{$\sim 0.0027$} \\
\hline
\end{tabular}
\end{table}

\subsubsection*{Pulse parameters}
All single-qutrit gates are implemented using Gaussian pulses with a duration of $71.11$ ns and a standard deviation $\sigma = 17.77$ ns with DRAG components. For qutrit 1, the $X^{01}$ gate uses an amplitude of $0.0211$ and a drag coefficient $\beta = -1.15$, while the $X^{12}$ gate uses an amplitude of $0.0145$ and $\beta = 0.38$. For qutrit 2, the $X^{01}$ gate has an amplitude of $0.0205$ and $\beta = -1.88$, and the $X^{12}$ gate uses an amplitude of $0.0145$ with $\beta = -0.97$.

The cross-resonance-based two-qutrit controlled-$X$, $U_{CX}^{01}$, is realized using a Gaussian Square envelope with a duration of $200$ ns, amplitude $\Omega (t)=1.1$ GHz, $\sigma = 16.89$ ns, and a square width of $180$ ns. Similarly, the controlled-$H$, $U_{CH}^{12}$, uses a shorter pulse with a duration of $100$ ns, amplitude $\Omega (t) =0.4$ GHz, $\sigma=16.89$ ns and a square width of $90$ ns. These pulse parameters are optimized to minimize leakage and phase errors across all transitions involved.



\subsection{Action of parametric gates}

\begin{figure}[h]
\centering
\includegraphics[width=0.5\linewidth]{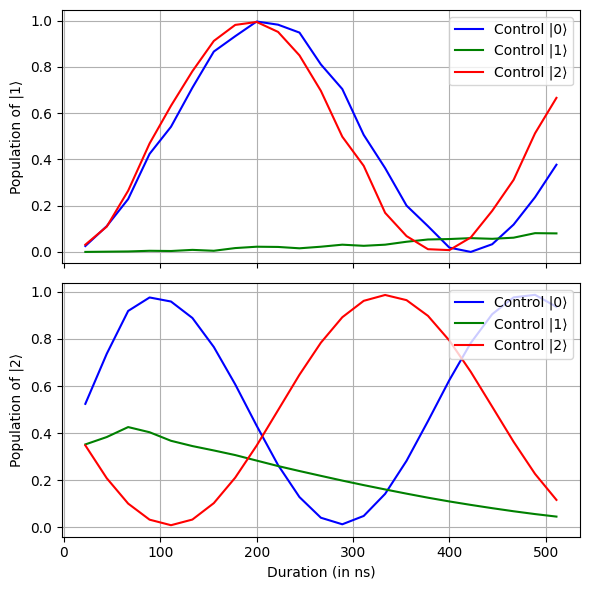}
\caption{The action of the proposed $\ucrzeroone$ and $\ucronetwo$ parametric gates over time.}
\label{fig:Fig2}
\end{figure}

The evolution of the target qutrit with an initial state $\ket{0}$ when $\ucrzeroone$ is applied. The frequency of the Rabi oscillations due to the control state $\ket{0}$ and the control state $\ket{2}$ differ very minimally. Our gate time parameters are essentially the same for both. The frequency of the Rabi oscillation due to the control state $\ket{1}$ is so small that the gate behaves as an identity operator when the control is $\ket{1}$ within the time of the gate. The evolution of the target qutrit with an initial state $\ket{\psi} = \ket{-}_{12} = \big(1/\sqrt{2}\big)(\ket{1}-\ket{2})$, when $\ucronetwo$ is applied to it. Similar to the $\ucrzeroone$, the gate acts almost as an identity when the control state is $\ket{1}$ at the gate time of $100$ ns. However, the Rabi oscillations induced when the control state is $\ket{0}$, and when it is $\ket{2}$, have the same frequency, but differ by a phase that is almost $\pi$, thereby inducing different rotations on the target state. We note that when the initial state is $\ket{1}$, the Rabi oscillations of both the control states $\ket{0}$ and $\ket{2}$ are the same.

\FloatBarrier
\subsection{Two-Qutrit Bell-State Preparation Circuit}

\begin{figure}[h]
\centering
\includegraphics[width=0.5\linewidth]{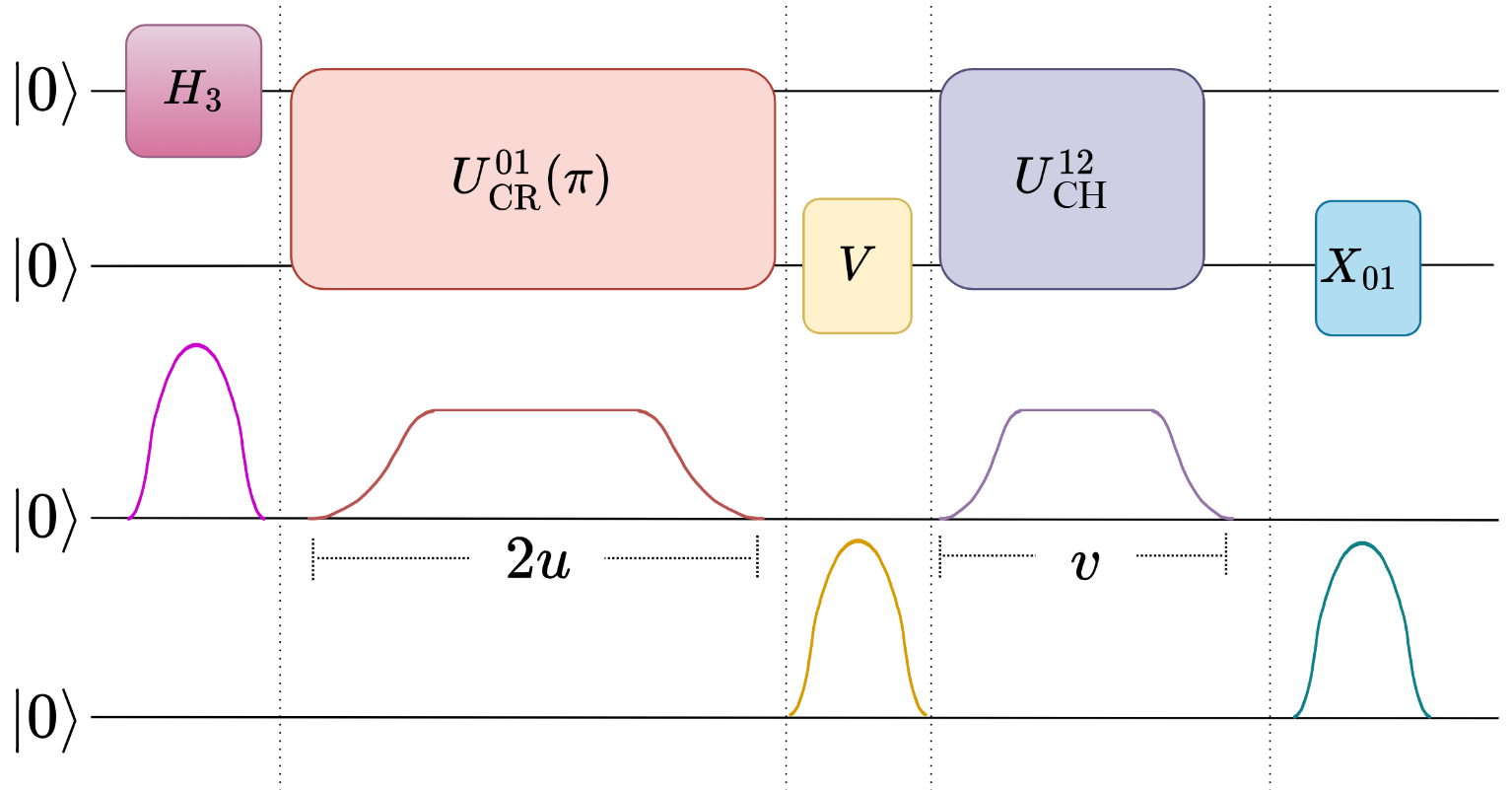}
\caption{The circuit diagrams corresponding to the two-qutrit Bell state preparation. $H_3$ denotes the qutrit Hadamard gate that maps the state |0⟩ to the equal superposition state $\ket{\psi} = (\ket{0} + \ket{1} + \ket{2})/\sqrt{3}$, $X_{01}$ denotes the single qutrit gate that performs the flip $\ket{0} \leftrightarrow \ket{1}$ and $V$ denotes the gate, $H_{12}$, that 
acts as the two-level Hadamard gate on the 1-2 level. For our choice of parameters, $2u = 200$ ns, $v = 100$ ns, and the total time for the preparation of the qutrit Bell state is $\sim 514$ ns.}
\label{fig:bell-circuit}
\end{figure}

\FloatBarrier
\subsection{Simulation Results}

\subsubsection{Bell-State complete QST}

\begin{figure}[ht]
\centering
\includegraphics[width=0.5\linewidth]{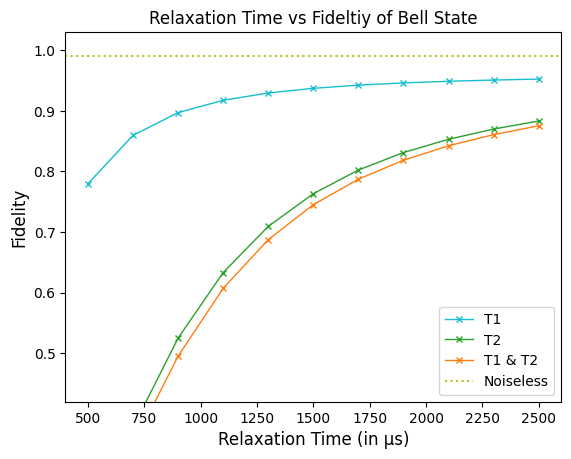}
\caption{The fidelity of the prepared Bell state over varying relaxation times. The dotted line corresponds to the fidelity of the Bell state in the noiseless setting. We notice that the effect of $T_2$ relaxation is significantly larger than that of $T_1$ relaxation for the same time values.}
\label{fig:Fig7}
\end{figure}

\FloatBarrier
\subsubsection{The QPT Plots of the proposed gates}

In this section, we present the QPT plots of the gates $\gcxzero, \gcxone, \gcsqrtzero$ and $\gcsqrtone$ as defined in the main manuscript, compared against the QPT plots of simulations of these gates in the noiseless and the noisy settings.

\begin{figure}[ht]
\centering
\includegraphics[width=0.95\linewidth]{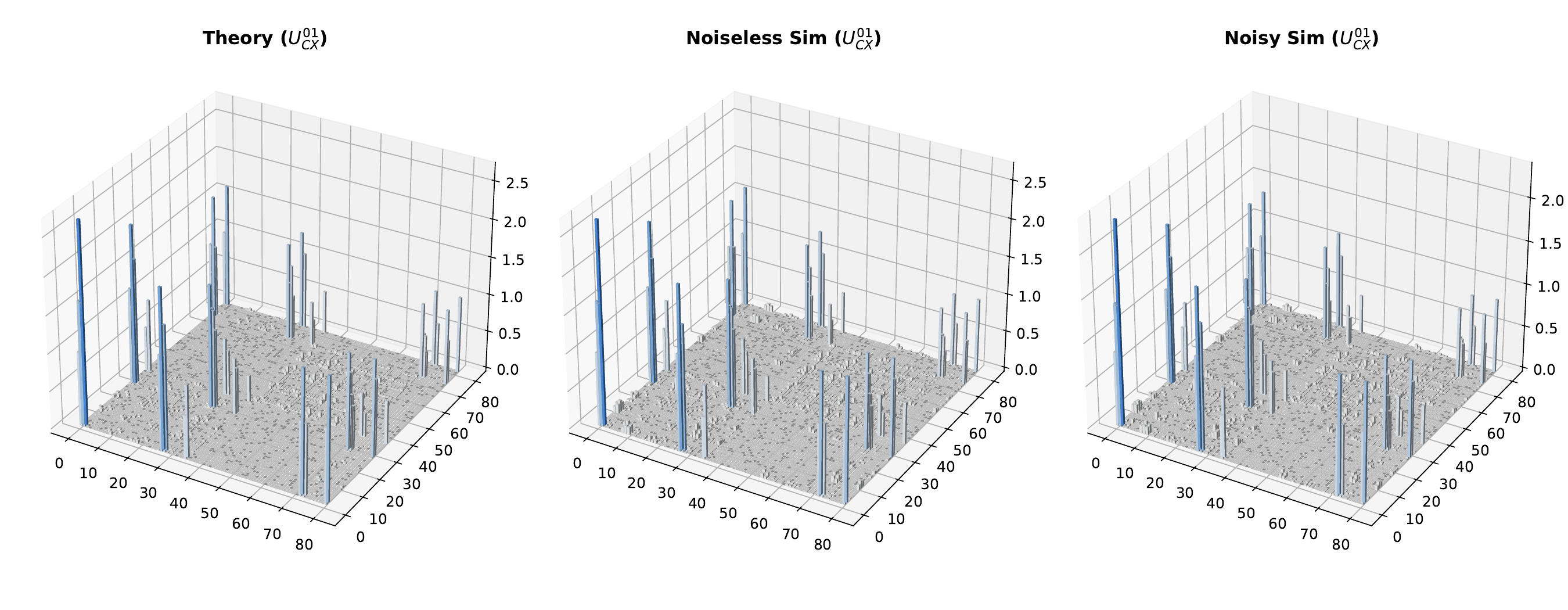}
\caption{Comparison between $\chi_{theory}$, $\chi_{noiseless}$ and $\chi_{noisy}$ obtained using QPT for the $\gcxzero$ gate.}
\label{fig:ucx01-qpt}
\end{figure}
\begin{figure}[ht]
\centering
\includegraphics[width=0.95\linewidth]{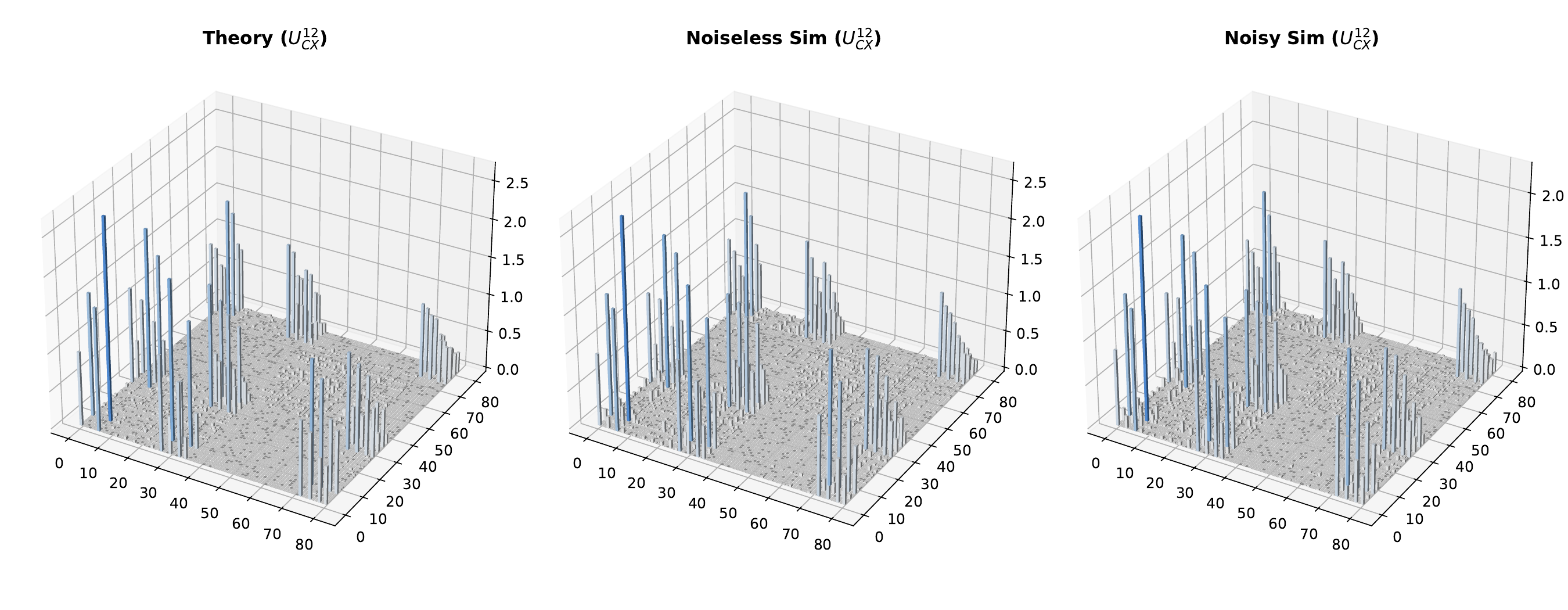}
\caption{Comparison between $\chi_{theory}$, $\chi_{noiseless}$ and $\chi_{noisy}$ obtained using QPT for the $\gcxone$ gate.}
\label{fig:ucx12-qpt}
\end{figure}
\begin{figure}[ht]
\centering
\includegraphics[width=0.95\linewidth]{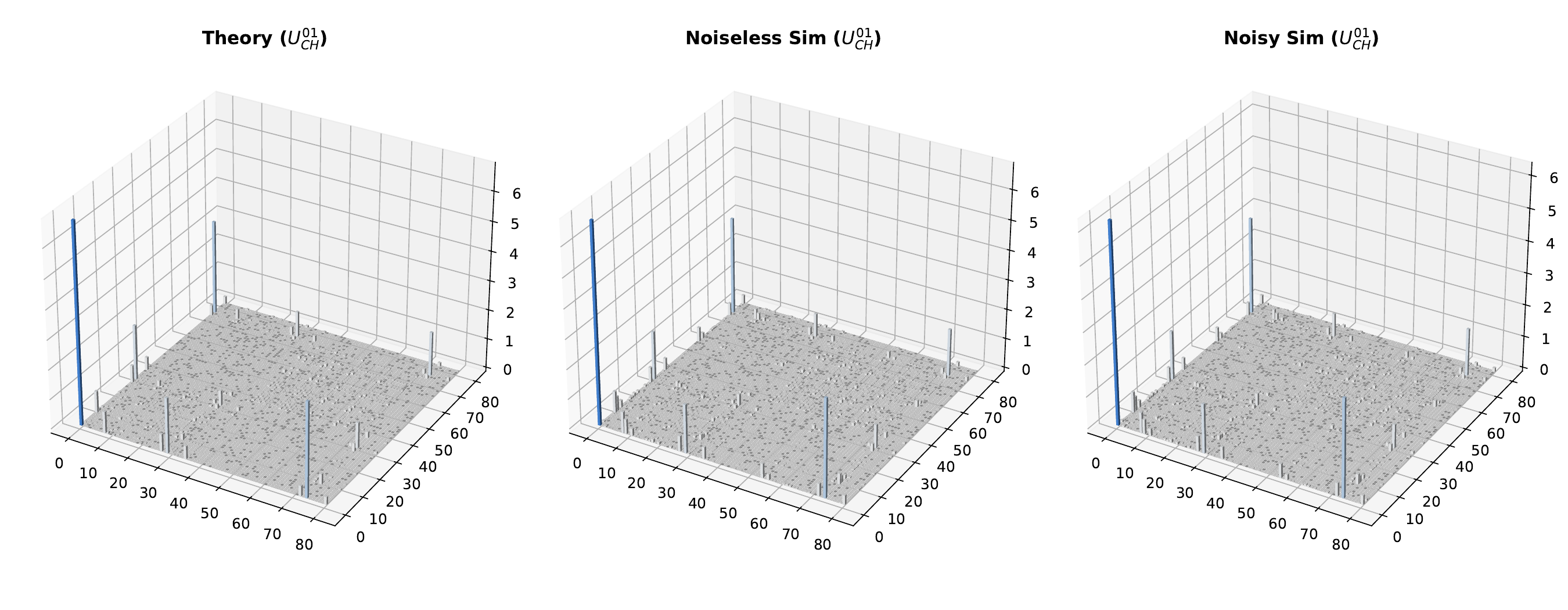}
\caption{Comparison between $\chi_{theory}$, $\chi_{noiseless}$ and $\chi_{noisy}$ obtained using QPT for the $\gcsqrtzero$ gate.}
\label{fig:uch01-qpt}
\end{figure}

\begin{figure}[ht]
\centering
\includegraphics[width=0.95\linewidth]{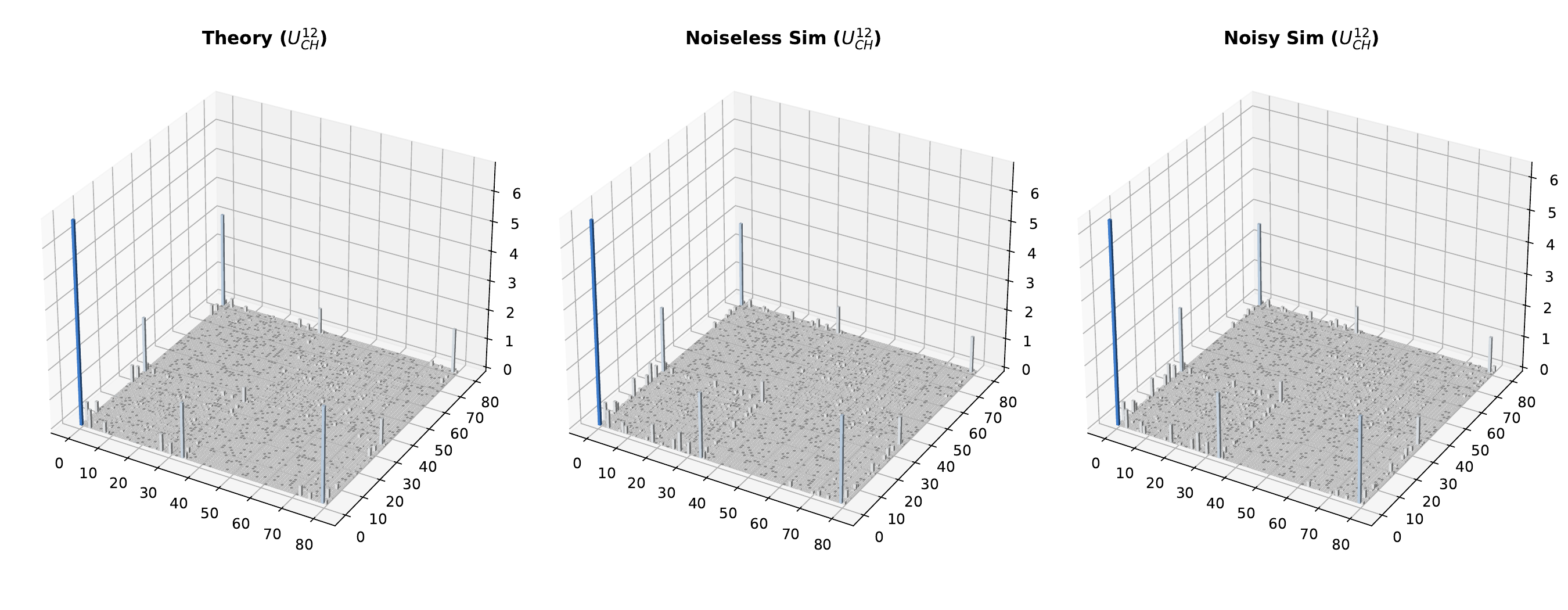}
\caption{Comparison between $\chi_{theory}$, $\chi_{noiseless}$ and $\chi_{noisy}$ obtained using QPT for the $\gcsqrtone$ gate.}
\label{fig:uch12-qpt}
\end{figure}

\FloatBarrier
\subsubsection{The Fidelity of Proposed Gates under Varying $T_1$ and $T_2$ Relaxation}

\begin{table*}[h]
\centering

\label{tab:gate_fidelities_percent}
\setlength{\tabcolsep}{4pt}
\renewcommand{\arraystretch}{1.15}

\begin{tabular}{c | c | *{11}{c}}
\toprule
\textbf{Gate} & \textbf{Noise} &
500 & 700 & 900 & 1100 & 1300 & 1500 & 1700 & 1900 & 2100 & 2300 & 2500 \\
\midrule

\multirow{3}{*}{$U_{\mathrm{CX}}^{01}$}
 & T1       & 84.68 & 91.56 & 94.66 & 96.29 & 97.25 & 97.86 & 98.27 & 98.56 & 98.77 & 98.93 & 99.05 \\
 & T2       & 25.12 & 47.40 & 62.99 & 73.09 & 79.73 & 84.24 & 87.42 & 89.74 & 91.46 & 92.78 & 93.81 \\
 & T1 \& T2 & 21.41 & 43.56 & 59.80 & 70.57 & 77.75 & 82.66 & 86.14 & 88.68 & 90.58 & 92.04 & 93.17 \\

\midrule

\multirow{3}{*}{$U_{\mathrm{CX}}^{12}$}
 & T1       & 83.39 & 89.98 & 92.96 & 94.54 & 95.47 & 96.06 & 96.46 & 96.74 & 96.94 & 97.10 & 97.22 \\
 & T2       & 25.20 & 46.85 & 62.01 & 71.85 & 78.33 & 82.74 & 85.84 & 88.10 & 89.79 & 91.08 & 92.09 \\
 & T1 \& T2 & 21.57 & 43.11 & 58.91 & 69.41 & 76.41 & 81.20 & 84.60 & 87.08 & 88.93 & 90.35 & 91.47 \\

\midrule

\multirow{3}{*}{$U_{\mathrm{CH}}^{01}$}
 & T1       & 91.41 & 95.19 & 96.81 & 97.65 & 98.14 & 98.45 & 98.66 & 98.81 & 98.91 & 98.99 & 99.05 \\
 & T2       & 46.83 & 66.09 & 76.60 & 82.65 & 86.38 & 88.81 & 90.48 & 91.68 & 92.56 & 93.23 & 93.74 \\
 & T1 \& T2 & 44.24 & 65.04 & 76.69 & 83.48 & 87.69 & 90.45 & 92.35 & 93.71 & 94.71 & 95.47 & 96.06 \\

\midrule

\multirow{3}{*}{$U_{\mathrm{CH}}^{12}$}
 & T1       & 91.09 & 94.82 & 96.43 & 97.26 & 97.75 & 98.05 & 98.26 & 98.41 & 98.51 & 98.59 & 98.65 \\
 & T2       & 47.50 & 66.25 & 76.58 & 82.56 & 86.25 & 88.67 & 90.33 & 91.52 & 92.39 & 93.06 & 93.57 \\
 & T1 \& T2 & 44.80 & 65.08 & 76.52 & 83.22 & 87.38 & 90.11 & 91.99 & 93.34 & 94.34 & 95.09 & 95.68 \\

\midrule

\multirow{3}{*}{Identity (200 ns)}
 & T1       & 85.56 & 92.25 & 95.30 & 96.92 & 97.87 & 98.48 & 98.89 & 99.18 & 99.39 & 99.55 & 99.67 \\
 & T2       & 32.47 & 51.13 & 65.05 & 74.42 & 80.72 & 85.07 & 88.16 & 90.42 & 92.12 & 93.43 & 94.45 \\
 & T1 \& T2 & 27.24 & 46.89 & 61.74 & 71.86 & 78.72 & 83.48 & 86.87 & 89.36 & 91.24 & 92.68 & 93.80 \\

\midrule

\multirow{3}{*}{Identity (100 ns)}
 & T1       & 92.49 & 96.24 & 97.86 & 98.71 & 99.20 & 99.51 & 99.72 & 99.86 & 99.97 & 100.00 & 100.00 \\
 & T2       & 51.77 & 69.80 & 80.08 & 86.12 & 89.89 & 92.37 & 94.08 & 95.31 & 96.21 & 96.90 & 97.44 \\
 & T1 \& T2 & 47.55 & 66.85 & 78.01 & 84.62 & 88.76 & 91.50 & 93.39 & 94.75 & 95.75 & 96.51 & 97.10 \\

\bottomrule
\end{tabular}

\caption{Gate fidelities (in percent) under different noise models as a function of relaxation time (in $\mu$s). The fidelity values are within an error tolerance of $\pm 0.01\%$}
\end{table*}

\FloatBarrier
 
\subsubsection{Fidelity Plots}

\begin{figure*}[h]
\centering
\begin{subfigure}[b]{0.42\textwidth}
\fbox{
   \includegraphics[width=\linewidth]{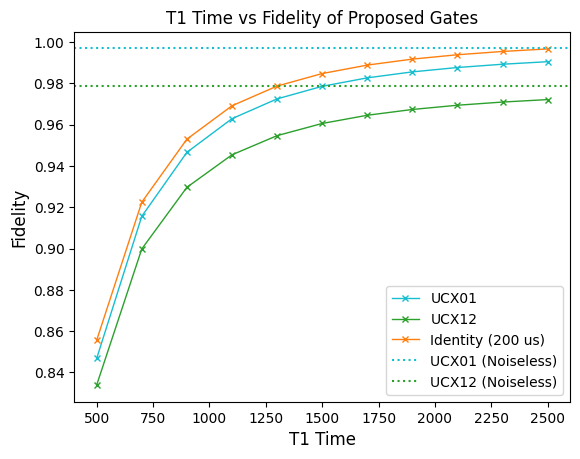}
   }
   \caption{}
\end{subfigure}
\hspace{0.5cm}
\begin{subfigure}[b]{0.42\textwidth}
    \fbox{
   \includegraphics[width=\linewidth]{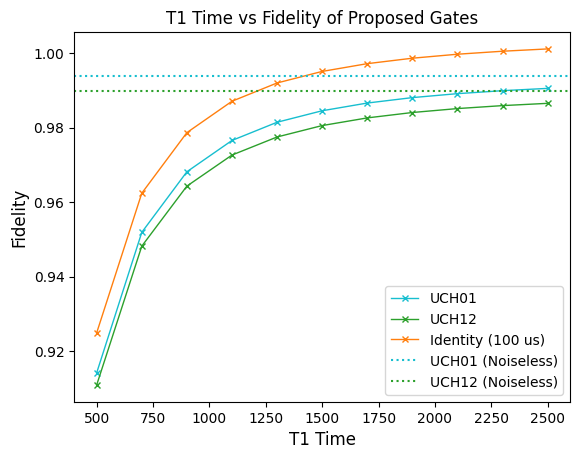}
   }
   \caption{}
\end{subfigure}

\captionsetup{justification=raggedright,singlelinecheck=false}

\caption{The T1 relaxation time (in $\mu s$) vs Fidelity Plots of the Proposed Gates}
\label{fig:fidelity-plots}
\end{figure*}

{
\clearpage
\raggedbottom
\begin{figure}[h]
\centering
\begin{subfigure}[b]{0.42\textwidth}
\fbox{
   \includegraphics[width=0.95\linewidth]{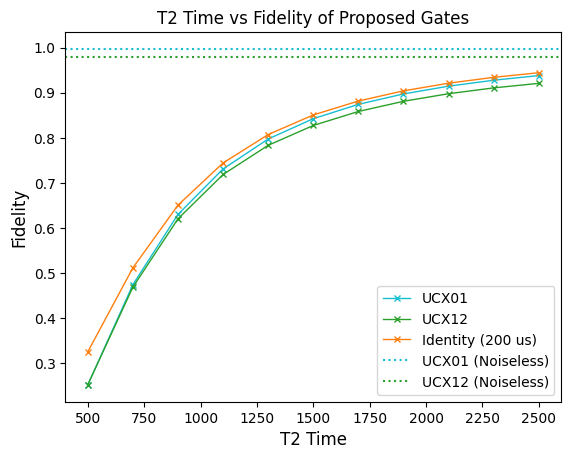}
   }
   \caption{}
\end{subfigure}
\hspace{0.5cm}
\begin{subfigure}[b]{0.42\textwidth}
    \fbox{
   \includegraphics[width=0.95\linewidth]{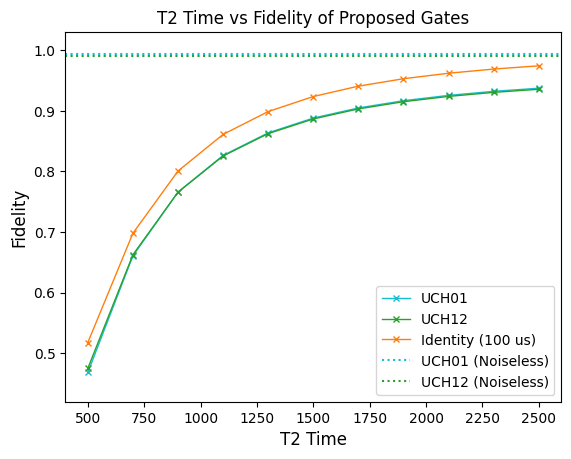}
   }
   \caption{}
\end{subfigure}

\captionsetup{justification=raggedright,singlelinecheck=false}

\caption{The T2 relaxation time (in $\mu s$) vs Fidelity Plots of the Proposed Gates}
\label{fig:t2-fidelity-plots}
\end{figure}

\begin{figure}[h]
\centering
\begin{subfigure}[b]{0.42\textwidth}
\fbox{
   \includegraphics[width=0.95\linewidth]{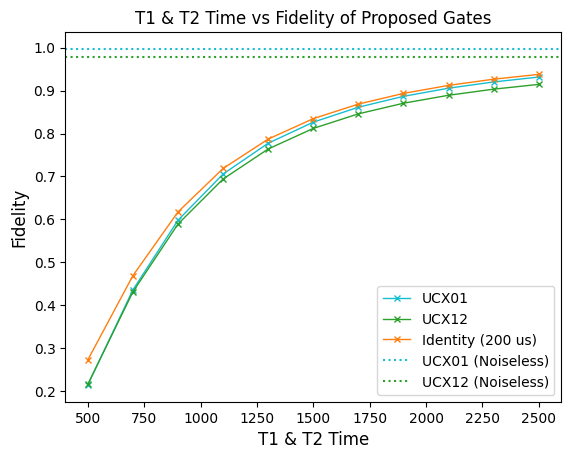}
   }
   \caption{}
\end{subfigure}
\hspace{0.5cm}
\begin{subfigure}[b]{0.42\textwidth}
    \fbox{
   \includegraphics[width=\linewidth]{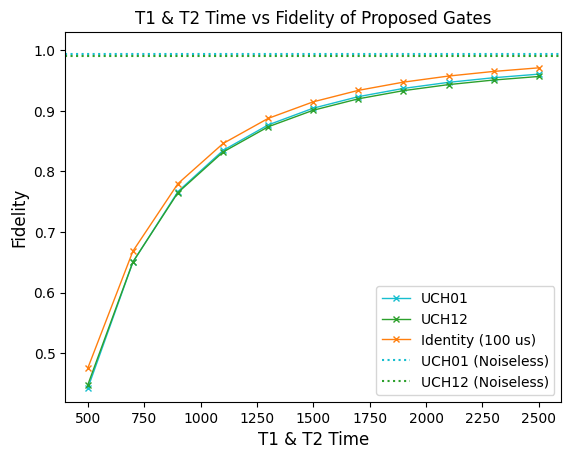}
   }
   \caption{}
\end{subfigure}

\captionsetup{justification=raggedright,singlelinecheck=false}

\caption{The T1 \& T2 relaxation time (in $\mu s$) vs Fidelity Plots of the Proposed Gates}
\label{fig:fidelity-plots}
\end{figure}
}

\end{document}